 \documentclass[twocolumn,pra,showpacs,floatfix]{revtex4-1}
\usepackage{graphicx}
\usepackage{subfigure}
\usepackage[]{amsmath}
\usepackage{dcolumn}
\usepackage{bm}
\usepackage{tabularx}
\RequirePackage{rotating}

\graphicspath{.}

\newcommand{\la}{\left\langle}
\newcommand{\ra}{\right\rangle}
\newcommand{\eg}{\emph{e.g.}\ }
\newcommand{\ie}{\emph{i.e.}\ }
\newcommand{\etal}{\emph{et al.}\ }
\newcommand{\EA}{^e\!\!A}

\newcommand{\beq}{\begin{equation}}
\newcommand{\eeq}{\end{equation}}
\newcommand{\beqa}{\begin{eqnarray}}
\newcommand{\eeqa}{\end{eqnarray}}
\newcommand{\beqn}{\begin{equation*}}
\newcommand{\eeqn}{\end{equation*}}
\newcommand{\beqan}{\begin{eqnarray*}}
\newcommand{\eeqan}{\end{eqnarray*}}

\newcommand{\ar}{\begin{array}}
\newcommand{\ear}{\end{array}}
\newcommand{\bc}{\begin{color}}
\newcommand{\ec}{\end{color}}
\newcommand{\bit}{\begin{itemize}}
\newcommand{\eit}{\end{itemize}}

\begin{document}

\title[A theoretical study of the C$^-$ bound states and C lowest configuration]{A theoretical study of the C$^-$ $^4S^o_{3/2}$ and $^2D^o_{3/2,5/2}$ bound states and C ground configuration: fine and hyperfine structures, isotope shifts and transition probabilities}

\author{T. Carette}\email{tcarette@ulb.ac.be}
\author{M. R. Godefroid}\email{mrgodef@ulb.ac.be}

\affiliation{Chimie quantique et photophysique, CP160/09, Universit\'e Libre de Bruxelles, B 1050 Brussels, Belgium}

\date{\today}

\begin{abstract}
This work is an \emph{ab initio} study of the $2p^3~^4S^o_{3/2}$, and $^2D^o_{3/2,5/2}$ states of C$^-$ and $2p^2~^3P_{0,1,2}$, $^1D_2$, and $^1S_0$ states of neutral carbon. We use the multi-configuration Hartree-Fock approach, focusing on the accuracy of the wave function itself.
We obtain all C$^-$ detachment thresholds, including correlation effects to about 0.5\%. 
Isotope shifts and hyperfine structures are calculated. The achieved accuracy of the latter is of the order of 0.1~MHz. Intra-configuration transition probabilities are also estimated.
\end{abstract}




\maketitle

\section{Introduction}

Negative ions have always attracted broad attention from the scientific community~\cite{Peg:04a,And:04a}. They challenge both the experimentalist and theoreticians. The first because they are weakly bound, and therefore fragile, and because they do not possess a lot of features allowing measurements. The latter because the binding of an extra electron is granted only by a arrangement of the electrons in a highly correlated system~\cite{Kraetal:87a}. Moreover, the fact that the electrons in negative ions are bound by a short range potential confer them unique properties.

\begin{figure}
\includegraphics[width=0.47\textwidth]{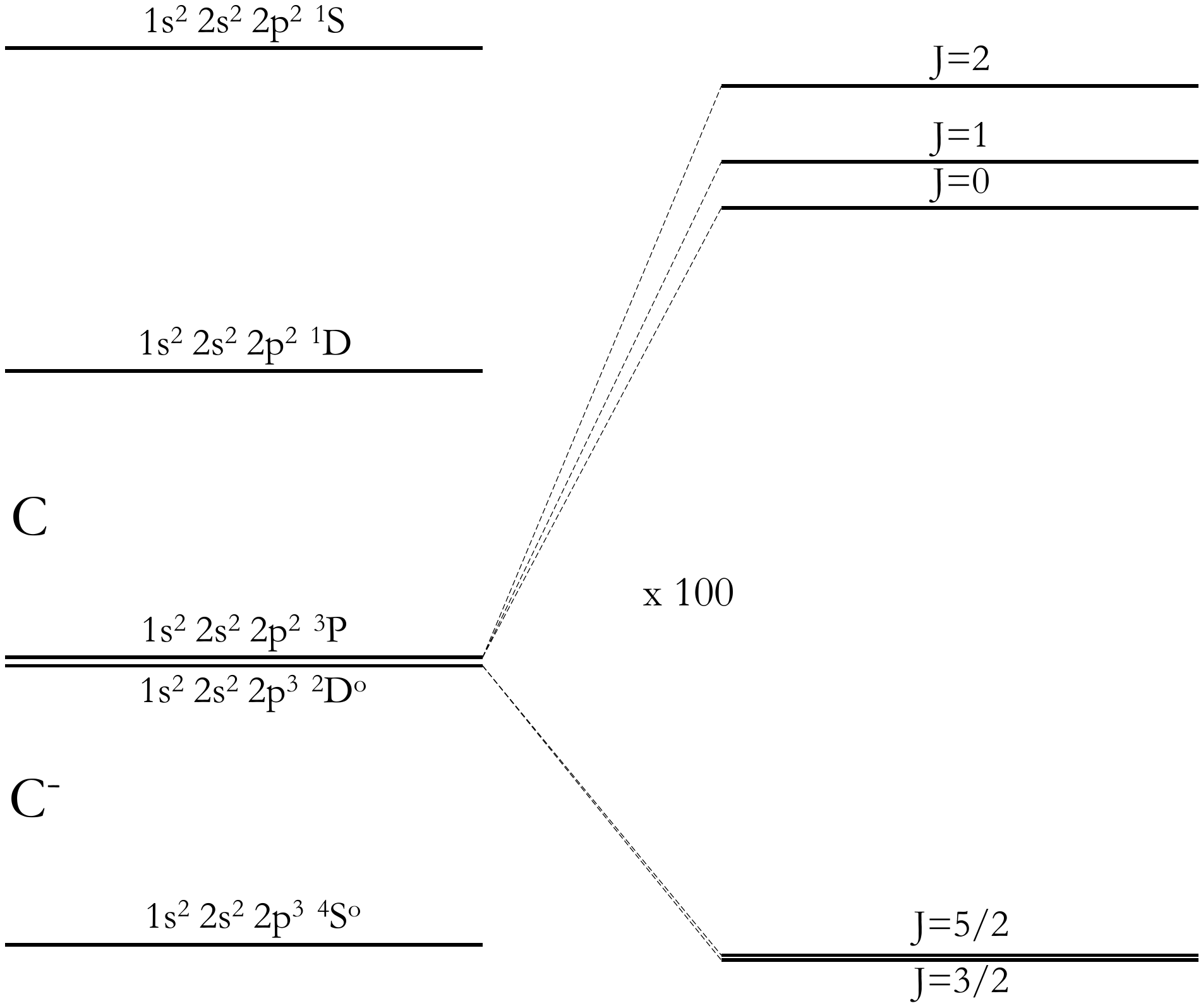}
\caption{Level diagram of the C and C$^-$. The fine structure of the $^3P$ state of the neutral carbon and the $^2D^o$ state of C$^-$ are magnified ($\times100$). \label{fig7:CLev}}

\end{figure}

C$^-$ is the lightest negative ion to have two bound terms: its ground state $^4S^o$ and the $^2D^o$ excited state which both arise from the $2p^3$ configuration. The level diagram of the states studied in this work is given in Figure~\ref{fig7:CLev}.

Carbon is among the most abundant components in the universe and a key element in life chemistry.
The carbon negative ion is attracting in astrophysics and atmosphere physics since nitrogen-like $2p^3~^4S^o -\, {^2D^o}$ forbidden lines are recognized as useful transitions for abundances determination~\cite{NemGod:09a,Shaetal:05a,Shaetal:08a}. It has also recently been suggested by \mbox{Le Padellec~\etal \cite{Padetal:08a}} that C$^-$ negative ion could intervene in astrophysical reactions. The photodetachment cross-sections of the C$^-$ have been repeatedly studied, both theoretically~\cite{Zhoetal:05a} and experimentally~\cite{Braetal:98a}, for photon energies addressing valence electrons and core electrons~\cite{KasIva:06a,Waletal:06a}. Recently, an isotope separation method was tested by Andersson \etal \cite{Andetal:08a}, based on the isotopic dependence of the Doppler shift of the C$^-$ detachment thresholds in an accelerator.

\medskip

A binding energy of $1.262~119(20)$ eV for the C$^-(^4S^o)$ has been measured by Scheer \etal \cite{Schetal:98a} who could not improve the old value of $33(1)$ meV for the C$^-(^2D^o)$ detachment threshold, measured by Feldmann~\cite{Fel:77a}. The fine-structure of the $^2D^o$ state is not known. On the theoretical side, very accurate carbon electron affinities were obtained with coupled-cluster-based methods~\cite{Olietal:99a,Kloetal:10a,Cleetal:11a}.

The structure of the C$^-$ has not been studied thoroughly and especially little is known about the $^2D^o$ multiplet. In laboratory plasmas, lifetimes of the order of the ms were measured for the C$^-(^2D^o)$, the electron detachment being principally caused by the black-body radiation and, to a lesser extent, to collisions~\cite{Taketal:07a}. Significantly longer lifetimes could be reached in the cold and diluted interstellar media where molecular anions have already been detected~\cite{Thaetal:08a}. However, the C$^-(^2D^o)$ is, for various reasons, very difficult to study experimentally. In this context, a firm theoretical knowledge of this system is particularly precious.

\medskip

Elements from boron to fluorine are the next targets after beryllium in the working line of ``exact" calculations. High accuracy can be achieved for systems with up to four electrons using wave functions expanded in explicitly correlated gaussian or in Hylleraas coordinates~\cite{Hyl:29a,Kometal:95a,Staetal:09a}. Although the precision that can be achieved for atoms with more electrons is limited by the complexity of the electron correlation mathematical treatment, the ground states of the second period $p-$block atoms from B to F are satisfactorily described by a non-relativistic approach on top of which relativistic corrections are added.

A critical benchmark quantity for highly correlated models is the isotope shift ($IS$) on the electron affinity ($\EA$) that is doubly sensitive to correlation effects: through the negative ion structure and through the specific mass shift parameter. The multi-configuration Hartree-Fock method has been successfully used for calculating the $IS$ on the $\EA$ of O~\cite{GodFro:99a}, S~\cite{Caretal:10a} and Cl~\cite{CarGod:11a}.

The purpose of the present work is double. Our principal objective is to obtain the crucial informations about the C$^-$ electronic structure for stimulating experimental research on the $^2D^o$ state. Therefore, we focus on quantities that are especially difficult for experimentalists to measure: isotope shifts, hyperfine structures and absolute transition probabilities. As for the energy separations themselves, we do not try to compete nor with the observation, nor with the previous coupled-cluster calculations. We instead use these reliable reference data for assessing the quality of our computational procedure.

Our second objective is to obtain non-relativistic (NR) wave functions as accurate as possible using the standard tools of the ATSP2K package~\cite{Froetal:07a}.
For getting the best estimation of the accuracy, we choose to use the same systematical construction of our C and C$^-$ models, avoiding any arbitrary compensation of the ``additional'' electron correlation of the negative ion compared to the neutral atom. 

\medskip

In Section~\ref{sec7:comp}, we present large scale numerical multi-configuration Hartree-Fock (MCHF) calculations (Section~\ref{sec7:nrcalc}), relativistic calculations estimated using the Breit-Pauli approach (BPCI, Section~\ref{sec7:bpci})~\cite{Froetal:97a} and relativistic configuration interaction based on the Dirac equation (RCI, Section~\ref{sec7:rci})~\cite{Jonetal:07a}. 
In Section~\ref{sec7:expth}, we present accurate results for hyperfine structures (Section~\ref{sec7:expthHFS}), total energies including the fine structure (Section~\ref{sec7:endiff}), mass polarization shift parameters (Section~\ref{sec7:ms}) of C~$2p^2\ ^3P$, $^1D$, $^1S$ and C$^-$~$2p^3\ ^4S^o$, $^2D^o$. In Section~\ref{sec7:trans}, we present the M1 and E2 transition probabilities within the $2p^2$ and $2p^3$ configurations of C and C$^-$.

\section{Computational method}\label{sec7:comp}

\subsection{The MCHF expansion}

The multiconfiguration Hartree-Fock (MCHF) variational approach consists in optimizing the 
one-electron functions spanning a configuration space and the mixing coefficients of the interacting configuration state functions (CSF)~\cite{Froetal:97a} for describing a given term
\begin{equation}
\label{eq:SOC}
\Psi(\gamma LS M_L M_S) = \sum_{i} c_i \Phi(\gamma_i LSM_L M_S).
\end{equation} 

\subsection{Hyperfine interaction}\label{sec3:HFS}

The level hyperfine structure is caused by the interaction of the angular momentum of the electrons ($\textbf{J}$) and of the nucleus ($\textbf{I}$), forming the total atomic angular momentum $\textbf{F}= \textbf{I} + \textbf{J}$.
The theory underlying the computation of hyperfine structure using MCHF wave functions can be found in references~\cite{LinRos:74a,Hib:75b,Jonetal:93a}. 
It is possible to express the non relativistic hyperfine interaction in terms of the $J$-independent orbital~($a_{l}$),
spin-dipole~($a_{sd}$), contact~($a_{c}$) and electric quadrupole ($b_{q}$) electronic hyperfine parameters defined as~\citep{LinRos:74a}
\begin{align}
\label{eq3:small_a_l}
a_{l}&\equiv\langle \Gamma LSM_{L}M_{S}|\sum_{i=1}^{N}
l^{(1)}_{0}(i)r^{-3}_{i}|\Gamma LSM_{L}M_{S}\rangle \; ,  \\
\label{eq3:small_a_sd}
a_{sd}&\equiv\langle \Gamma LSM_{L}M_{S}|\sum_{i=1}^{N}
2C^{(2)}_{0}(i)s^{(1)}_{0}(i)r^{-3}_{i}|\Gamma LSM_{L}M_{S}\rangle  \; , \\
\label{eq3:small_a_c}
a_{c}&\equiv\langle \Gamma LSM_{L}M_{S}|\sum_{i=1}^{N}
2s^{(1)}_{0}(i)r^{-2}_{i}\delta(r_{i})|\Gamma LSM_{L}M_{S}\rangle  \; , \\
\label{eq3:small_b_hfs_par}
b_{q} &\equiv\langle \Gamma LSM_{L}M_{S}|\sum_{i=1}^{N}
2C^{(2)}_{0}(i)r^{-3}_{i}|\Gamma LSM_{L}M_{S}\rangle  \; ,
\end{align}
and calculated for the magnetic component $M_{L}=L$ and $M_{S}=S$~\citep{Hib:75a}. The diagonal hyperfine interaction energy correction is usually expressed in terms of the hyperfine magnetic dipole ($A_J$) and electric quadrupole ($B_J$) constants as follows
\beqa
W(J,J) &=& A_J\frac{C}{2} \nonumber\\
&&+ B_J \frac{3C(C+1) - 4 I(I+1)J(J+1)}{8I(2I-1)J(2J-1)}.
\eeqa
The first three parameters~(\ref{eq3:small_a_l}),~(\ref{eq3:small_a_sd}), and~(\ref{eq3:small_a_c}) contribute to the magnetic dipole hyperfine interaction constant through
\begin{equation}
\label{A_constant}
A_J = A_{J}^{l} + A^{sd}_{J} + A^{c}_{J} \; ,
\end{equation}
with~\cite{Jonetal:10a}
\begin{eqnarray}
\label{sec3:A_orb}
  A_{J}^{l} & = & G_{\mu}\frac{\mu_{I}}{I}\;a_{l} \; \frac{\langle\,\pmb{L}\cdot\pmb{J}\,\rangle}{LJ(J+1)}  \; ,\\
 \label{sec3:A_sd}
A^{sd}_{J} &= &\frac{1}{2}\,G_{\mu}\,g_{s}\,\frac{\mu_{I}}{I}\,a_{sd} \; \nonumber\\ &&\hspace{-0.5cm}\times\frac{3\,\langle\,\pmb{L}\cdot\pmb{S}
\,\rangle\,
\langle\,\pmb{L}\cdot\pmb{J}\,\rangle\,-\,L(L+1)\,\langle\,\pmb{S}\cdot\pmb{J}\,\rangle}{SL(2L-1)J(J+1)} \; ,\\
\label{sec3:A_c}
A_{J}^{c} &=&\frac{1}{6}\,G_{\mu}\,g_{s}\,\frac{\mu_{I}}{I}a_{c} \;
\frac{\langle\,\pmb{S}\cdot\pmb{J}\,\rangle}{SJ(J+1)} \; ,
\end{eqnarray}
while the last one ($b_q$) constitutes the electronic contribution to the electric quadrupole hyperfine interaction
\begin{eqnarray}
\label{sec3:B_constant}
B_{J}&&=-G_{q}\,Q\,b_{q}\,\nonumber\\ 
&&\times\frac{6\langle\,\pmb{L}\cdot\pmb{J}\,\rangle^{2}\,-\,3\langle\,\pmb{L}\cdot\pmb{J}\,
\rangle\,-\,2L(L+1)J(J+1)}{L(2L-1)(J+1)(2J+3)} \; .
\end{eqnarray}
Expressing the electronic parameters $a_l$, $a_{sd}$ and $a_c$ in atomic units ($a_0^{-3}$) and $\mu_I$ in nuclear magnetons ($\mu_N$), the magnetic dipole hyperfine structure constants $A_J$ are calculated in units of frequency (MHz) by using $G_\mu = 95.41067$. Similarly, the  electric quadrupole hyperfine structure constants $B_J$ are expressed in MHz  when adopting atomic units ($a_0^{-3}$) for $b_q$, barns for $Q$ and $G_q = 234.96475$. The expectation values of the angular momenta scalar products are given by
\begin{align}
\la \pmb{L}\cdot \pmb{J}\ra &= [J(J+1) + L(L+1) - S(S+1) ]/2, \\
\la \pmb{S}\cdot \pmb{J}\ra &= [J(J+1) - L(L+1) + S(S+1) ]/2, \\
\la \pmb{S}\cdot \pmb{L}\ra &= [J(J+1) - L(L+1) - S(S+1) ]/2 .
\end{align}
when calculated with non-relativistic $LSJ$ wave functions.
The expression for the off-diagonal hyperfine interaction, depending on the hyperfine constants $A_{J,J-1}$, $B_{J,J-1}$ and $B_{J,J-2}$, are developed in reference~\cite{Jonetal:93a}. Hibbert~\cite{Hib:75a} gives the expressions of $A_{J,J-1}$ in terms of the hyperfine parameters~(\ref{eq3:small_a_l}--\ref{eq3:small_b_hfs_par}).

\subsection{The isotope shift}

The first order isotope shift on an energy level is decomposed in a field shift (or volume shift) and a mass shift~\cite{Kin:84a}. The first is proportional to the change in nucleus rms radius and change of the modified electron density at the origin. It is negligible in our context.

The energy corrected for the first order mass shift, on the other hand, can be estimated using~\cite{Godetal:01a}
\beq
E_M = \frac{M}{m+M} E_\infty + \frac{Mm}{(M+m)^2} \frac{\hbar^2}{m} S_{sms}
\eeq
where $m$ is the electron mass, $M$ is the bare nucleus mass, $E_\infty$ the infinite mass nucleus and
\beq
S_{sms} = -\la \Psi_\infty \left| \sum_{i<j} \bm{\nabla}_i \cdot \bm{\nabla}_j \right| \Psi_\infty \ra\ .
\eeq
The first term contains the normal mass shift
\beq
\text{NMS}=-\frac{m}{m+M} E_\infty
\eeq
and the second one is the specific mass shift (SMS). The mass polarization parameter, $S_{sms}$, has the dimension of an inverse square length.

\subsection{Transition probabilities}

The Einstein $A_{if}$ coefficient of spontaneous emission is  defined as the total probability per unit of time for an atom in a given energy level $i$  to make a transition to any state of the energy level $f$~\cite{Cow:81a}.

A transition between levels of same parity is forbidden in the electric dipole approximation, being in general many orders of magnitude lower than an allowed transition. Two interactions of the same order of magnitude can contribute to the appearance of such transitions: the dipole magnetic and the electric quadrupole radiation-matter interactions.
At the non-relativistic level, a dipole magnetic transition (M1) is governed by the electronic magnetic dipole operator that is
\beq
A^{\text{M1}} \propto (E_i-E_f)^3 \left| \la \Gamma_f J_f \left|\left| \pmb{L} + g_s \pmb{S} \right|\right| \Gamma_i J_i \ra\right|^2. \label{eq7:M1}
\eeq
In the mono-configuration approximation, the above matrix element is non-zero only between states of the same configuration and $LS$. This selection rule is relaxed by configuration and $LS$ mixings, the remaining constraints being that $J_f=J_i, J_i\pm 1$, and that $\Psi_i$ and $\Psi_f$ have the same parity. For its part, an electric quadrupole (E2) transition rate is proportional to the electric quadrupole moment matrix element
\beq
A^{\text{E2}} \propto (E_i-E_f)^5 \left| \la \Gamma_f J_f \left|\left| \sum_k r_k^2 \pmb{C}^{(2)}(k) \right|\right| \Gamma_i J_i \ra\right|^2\, , \label{eq7:E2}
\eeq
the sum running on all spatial electron coordinates $k$. Neglecting the $LS$ term mixing, a necessary condition for $A_\text{E2}$ to be non-zero is that $S_f-S_i=0$, $|L_f-L_i|\leq 2$, $|L_f+L_i|\geq 2$ and that the atomic parity is conserved.

\subsection{Non-relativistic calculations}\label{sec7:nrcalc}
We first select a zero-order set of CSFs, the multi-reference (MR). For all studied states it is the set of single and double excitations of the main configuration
to the $n=2,3$ shells. 
All the CSFs interacting to first order with the MR are selected and we choose the reverse order for the subshell coupling~\cite{Caretal:10a}. The orbital active set is defined as the set of all orbitals characterized by quantum numbers $n \leq n_{max}$ and $l\leq l_{max}$, and is denoted  $\lceil n_{max} l_{max} \rceil$. We first performed calculations defined in the spaces $\lceil 4f \rceil$ to $\lceil 12k \rceil$, denoted \mbox{MR-I$\lceil n_{max} l_{max}\rceil$}.

For each active space
$\lceil 10k\rceil$,
$\lceil 11k\rceil$ and $\lceil 12k\rceil$, we order the configurations according to their weight
\footnote{The weight of a configuration is defined as
\mbox{$
w= \left( \sum_i c_{i}^2 \right)^{1/2}
$}
where the sum runs over the CSFs belonging to the configuration.}.
We then construct several new MRs following this hierarchy, independently for each state and active set, by selecting the minimum group of configurations that add up to a certain percentage $p$ of the total wave function. Those multi-references are denoted MR$_p$ for each given MR-I wave function. Unsurprisingly, the MR$_p$ sets are not sensitive to the used active set.

\begin{figure}[h!]
\includegraphics[width=0.47\textwidth]{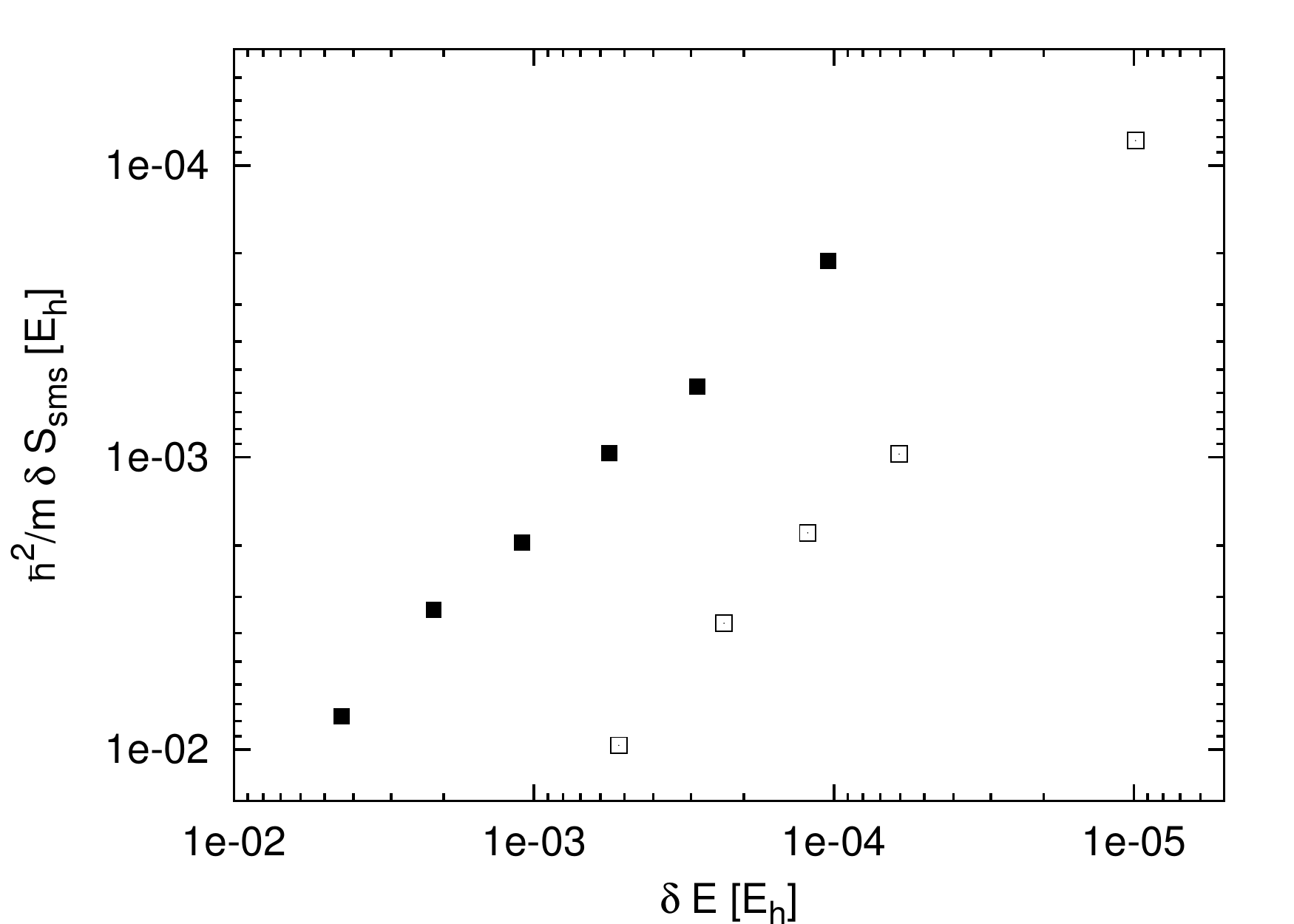}
\caption
{Bi-logarithmic plot of the convergence of the C($^3P$) mass polarization expectation value versus the corresponding energy, in Hartrees~(E$_h$). The coordinates of the black squares are the differences (in absolute value) between the results of the MR-I$\lceil nl \rceil$ and MR-I$\lceil 12k \rceil$ calculations, $n=6-11$ from left to right. Similarly, the coordinates of the white squares show the convergence of MR$_p$-I$\lceil 12k \rceil$ toward MR$_{99.95}$-I$\lceil 12k \rceil$, $p=(99-99.9)\%$ from left to right.\label{fig7:conv}}
\end{figure}

Configuration-interaction calculations are performed on each MR$_p$-I CSF sets, $p$ being limited to $99.8\%$ for C$^-(^4S^o)$ and to $99.3\%$ for C$^-(^2D^o)$. An example of the convergence of the calculations with the number of correlation layers ($n-2$) and $p$ is given in Figure~\ref{fig7:conv}. It presents results on the $2p^2~^3P$ state of neutral carbon. The black squares show $\frac{\hbar^2}{m}\delta S_{sms}$ with
\beqa
\delta S_{sms}=
|S_{sms}(\text{MR-I}\lceil nl \rceil) - S_{sms}(\text{MR-I}\lceil 12k \rceil)|,
\eeqa
versus
\beq
\delta E=|E(\text{MR-I}\lceil nl \rceil) - E(\text{MR-I}\lceil 12k \rceil)|
\eeq
for $n=6-11$. Similarly, the white squares compare the $E$ and $S_{sms}$ convergences of the MR$_{p}$-I$\lceil 12k \rceil$, $p=99.0-99.9$,  results toward the MR$_{99.95}$-I$\lceil 12k \rceil$ model. The energy always decreases along a sequence of increasingly large calculations, according to the variational principle, while the $S_{sms}$ of MR$_p$-I$\lceil nl \rceil$ calculations decreases with $n$ and increases with $p$. $\delta S_{sms}$ and $\delta E$ show a close to linear correlation, \ie the angular coefficients in the log-log figure is $\sim 1$. The slope of this correlation is slightly smaller than one for the convergence in $n$ and is about $10-20$ for the convergence in $p$, as can be seen from the offsets in the log-log figure. Similar behaviors where found in the open-core CI calculations of S/S$^-$~\cite{Caretal:10a} and Cl/Cl$^-$~\cite{CarGod:11a}. In general, we can make the following observations~\cite{Car:10a}: the CSFs that are important for the energy are accordingly important for the $S_{sms}$,
and the $S_{sms}$ value is more sensitive to the choice of the CSF space than to the orbital basis set.

The energy, $S_{sms}$ and hyperfine parameters calculated in this work are presented in Table~\ref{tab7:resNR_C} for neutral carbon and in Table~\ref{tab7:resNR_C-} for C$^-$.
It should be stressed that the differences between the results obtained with $p=99.9$ and $p=99.95$ are close to the expected numerical accuracy. We estimate that the dominant error in the neutral carbon calculations is due to the $nl$-truncation of the active set. This is not the case in the C$^-$ calculations for which the $p$-truncation is the most limiting.

From Tables~\ref{tab7:resNR_C} and~\ref{tab7:resNR_C-}, we observe that the differences between the MR$_p$-I$\lceil 11k \rceil$ and \mbox{MR$_p$-I$\lceil 12k \rceil$} results do not depend strongly on $p$. In fact, the error made by reporting the impact of the $12^\text{th}$ shell on the calculation with $p=99$, on the results of the largest MR$_p$-I$\lceil 11k\rceil$ is smaller than $4~10^{-7}$ E$_h$ on the energy, than $2~10^{-6}~a_0^{-2}$ on $S_{sms}$ and than $9~10^{-5}~a_0^{-3}$ on the hyperfine parameters.

Using this observation, we add a correction for the $13^\text{th}$ correlation layer and for the $l=8$ orbitals. First MCHF calculations are performed on the MR-I$\lceil 13k\rceil$ and MR-I$\lceil 12l\rceil$ CSF spaces, fixing all one-electron radial functions at the MR-I$\lceil 12k\rceil$ level and varying only the new orbitals. We use the so optimized orbitals in MR$_{99}$-I$\lceil 13l\rceil$ CI calculations, omitting the $13l$ subshell in the active set and using the multi-reference obtained with the $\lceil 12k\rceil$ active set. The two contributions (higher $n$ and $l$) are of same order of magnitude, as far as the energy is concerned. Still, the additional correlation layer tends to dominate in C$^-$ while the additional angular flexibility has the largest impact in the neutral carbon calculations.

As we already mentioned, the convergences in either $n$ or $p$, of a state energy and $S_{sms}$ are monotone and correlated (see \eg Figure~\ref{fig7:conv}). This fact could help strongly for extrapolating the energy and $S_{sms}$ value. However, even for a two-electron system, the precise behavior of the energy convergence with the principal quantum number in the high $n$-limit is unknown~\cite{Kut:08b,Heletal:08a}.
Froese-Fischer used the following extrapolation function for studying four electrons systems~\cite{Fro:93a}
\beq
\Delta E_n = a_4/(n-\delta n)^4 + a_5/(n-\delta n)^5 + a_6/(n-\delta n)^6
\label{eq7:extr}
\eeq
the $a_4$, $a_5$ and $a_6$ parameters, and $\delta n$ being chosen such that $a_4<0$ and $|a_4| \sim |a_5| \sim |a_6|$.
Fitting the $n=10-12$ results of Table~\ref{tab7:resNR_C} to equation (\ref{eq7:extr})
 for extrapolating to $n \rightarrow \infty$, we obtain $-37.84465$~E$_h$ for the energy of the C$(^3P)$ state. This procedure does not extrapolate to $l \rightarrow \infty$. The error on the extrapolation is of the order of $10^{-5}$~E$_h$ and the truncation in $l$ of about $10^{-4}$~E$_h$. We are in fair agreement with the non-relativistic energy of $-37.8450$ E$_h$ estimated by Chakravorty~\etal \cite{Chaetal:93a}. To our knowledge, the values of Table~\ref{tab7:resNR_C} are the best \emph{ab initio} estimated energies, even without the $n=13$ and $l=8$ corrections.
 Finally, let us mention that Sarsa~\etal \cite{Saretal:99a} calculated the $S_{sms}$ expectation value for the carbon $^3P$ state using the Monte Carlo (MC) approach with an explicitly correlated wave function and obtained $S_{sms}=-0.38(2)$~$a_0^{-2}$. Our final estimated value ($-0.40314~a_0^{-2}$) falls a bit outside the statistical MC error bars.

%
%

\begingroup \squeezetable
\begin{turnpage}
\begin{table*}
\caption{Results of the MCHF and CI calculations performed for the carbon $2p^2~^3P$, $^1D$ and $^1S$ terms. The energies $E$ are in E$_h$, the $S_{sms}$ in $a_0^{-2}$ and the hyperfine parameters in $a_0^{-3}$. The final values are the results of the larger $\lceil 12k\rceil$ calculations on which the impact of the $13^\text{th}$ shell and $l=8$ orbitals have been additively transferred. \label{tab7:resNR_C}}
\begin{ruledtabular}
\begin{tabular}{clcddddddcddddcdd}
\multicolumn{2}{c}{model} & \multicolumn{6}{c}{$1s^22s^22p^2~^3P$} & \multicolumn{4}{c}{$1s^22s^22p^2~^1D$} & \multicolumn{2}{c}{$1s^22s^22p^2~^1S$}\\
\cline{1-2}\cline{4-9}\cline{11-14}\cline{16-17}
$nl$ & \multicolumn{1}{c}{$p$} && \multicolumn{1}{c}{$E$} & \multicolumn{1}{c}{$S_{sms}$} & \multicolumn{1}{c}{$a_l$} & \multicolumn{1}{c}{$a_{sd}$} & \multicolumn{1}{c}{$a_c$} & \multicolumn{1}{c}{$b_q$} && \multicolumn{1}{c}{$E$} & \multicolumn{1}{c}{$S_{sms}$} & \multicolumn{1}{c}{$a_l$} & \multicolumn{1}{c}{$b_q$} && \multicolumn{1}{c}{$E$} & \multicolumn{1}{c}{$S_{sms}$} \\
\cline{1-2}\cline{4-9}\cline{11-14}\cline{16-17}\\
&&&\multicolumn{6}{c}{MCHF}&&\multicolumn{4}{c}{MCHF}&&\multicolumn{2}{c}{MCHF}\\
\\
HF     &            && -37.688618 & -1.39418 & 1.69181 & 0.33836 & 0.0     & 0.67672 && -37.631331 & -1.35557 & 3.26420 & -1.30568 && -37.549610 & -1.29745 \\
 $ 4 $ &            && -37.823094 & -0.43410 & 1.68577 & 0.37538 & 0.26753 & 0.60177 && -37.773728 & -0.40151 & 3.23862 & -1.14955 && -37.720089 & -0.38101 \\
 $ 5 $ &            && -37.834178 & -0.40791 & 1.70216 & 0.36515 & 0.59931 & 0.61023 && -37.786495 & -0.37677 & 3.27425 & -1.16232 && -37.733658 & -0.38198 \\
 $ 6 $ &            && -37.839793 & -0.39926 & 1.70450 & 0.35943 & 0.46199 & 0.63127 && -37.792593 & -0.36238 & 3.27661 & -1.20179 && -37.740170 & -0.35936 \\
 $ 7 $ &            && -37.842009 & -0.40361 & 1.70480 & 0.36087 & 0.42598 & 0.63566 && -37.795060 & -0.36697 & 3.27734 & -1.21478 && -37.742742 & -0.36255 \\
 $ 8k$ &            && -37.843075 & -0.40499 & 1.70463 & 0.36235 & 0.44763 & 0.63110 && -37.796285 & -0.36842 & 3.27694 & -1.20731 && -37.743993 & -0.36275 \\
 $ 9k$ &            && -37.843607 & -0.40598 & 1.70460 & 0.36107 & 0.48183 & 0.63025 && -37.796884 & -0.36938 & 3.27673 & -1.20173 && -37.744656 & -0.36358 \\
 $10k$ &            && -37.843885 & -0.40637 & 1.70460 & 0.36122 & 0.46815 & 0.63195 && -37.797231 & -0.36991 & 3.27669 & -1.20672 && -37.745021 & -0.36384 \\
 $11k$ &            && -37.844065 & -0.40673 & 1.70461 & 0.36136 & 0.47197 & 0.63174 && -37.797436 & -0.37029 & 3.27665 & -1.20611 && -37.745234 & -0.36417 \\
 $12k$ &            && -37.844170 & -0.40695 & 1.70462 & 0.36137 & 0.47401 & 0.63133 && -37.797556 & -0.37051 & 3.27155 & -1.20364 && -37.745361 & -0.36436 \\
\cline{1-2}\cline{4-9}\cline{11-14}\cline{16-17}\\
&&&\multicolumn{6}{c}{CI}&&\multicolumn{4}{c}{CI}&&\multicolumn{2}{c}{CI}\\
\\
 $11k$ & $99.0 $    && -37.843780 & -0.41241 & 1.70500 & 0.36141 & 0.47079 & 0.63192 && -37.797309 & -0.37136 & 3.27684 & -1.20610 && -37.744893 & -0.36693\\
       & $99.5 $    && -37.844069 & -0.40643 & 1.70495 & 0.36143 & 0.47529 & 0.63186 && -37.797568 & -0.36819 & 3.27677 & -1.20604 && -37.745357 & -0.36179\\
       & $99.7 $    && -37.844179 & -0.40455 & 1.70474 & 0.36139 & 0.46821 & 0.63181 && -37.797740 & -0.36500 & 3.27651 & -1.20576 && -37.745523 & -0.35869\\
       & $99.8 $    && -37.844255 & -0.40344 & 1.70510 & 0.36150 & 0.45312 & 0.63185 && -37.797794 & -0.36429 & 3.27653 & -1.20581 && -37.745561 & -0.35805\\
       & $99.9 $    && -37.844291 & -0.40282 & 1.70508 & 0.36151 & 0.45521 & 0.63183 && -37.797830 & -0.36388 & 3.27665 & -1.20586 && -37.745588 & -0.35766\\
       & $99.95$    && -37.844301 & -0.40274 & 1.70507 & 0.36151 & 0.45549 & 0.63182 && -37.797840 & -0.36374 & 3.27661 & -1.20583 && -37.745600 & -0.35750\\
\\
 $12k$ & $99.0 $    && -37.843884 & -0.41263 & 1.70501 & 0.36141 & 0.47288 & 0.63150 && -37.797430 & -0.37156 & 3.27675 & -1.20574 && -37.745020 & -0.36712\\
       & $99.3 $    &&            &          &         &         &         &         && -37.797596 & -0.36938 & 3.27692 & -1.20577 && -37.745388 & -0.36357\\
       & $99.5 $    && -37.844173 & -0.40664 & 1.70495 & 0.36143 & 0.47731 & 0.63145 && -37.797689 & -0.36838 & 3.27668 & -1.20567 && -37.745483 & -0.36198\\
       & $99.7 $    && -37.844284 & -0.40476 & 1.70474 & 0.36140 & 0.47023 & 0.63140 && -37.797860 & -0.36523 & 3.27644 & -1.20542 && -37.745655 & -0.35871\\
       & $99.8 $    && -37.844346 & -0.40392 & 1.70510 & 0.36150 & 0.45486 & 0.63145 && -37.797914 & -0.36452 & 3.27649 & -1.20550 && -37.745689 & -0.35821\\
       & $99.9 $    && -37.844396 & -0.40303 & 1.70509 & 0.36152 & 0.45670 & 0.63143 && -37.797951 & -0.36407 & 3.27661 & -1.20558 && -37.745716 & -0.35782\\
       & $99.95$    && -37.844406 & -0.40295 & 1.70507 & 0.36152 & 0.45698 & 0.63142 && -37.797962 & -0.36394 & 3.27657 & -1.20556 && -37.745728 & -0.35766\\
\\
 $13l$ & $99.0 $    && -37.844003 & -0.41282 & 1.70505 & 0.36137 & 0.47353 & 0.63153 && -37.797580 & -0.37182 & 3.27690 & -1.20549 && -37.745181 & -0.36736\\
\\
\multicolumn{2}{c}{Final} && -37.844525 & -0.40314 & 1.70511 & 0.36148 & 0.45699 & 0.63144 && -37.798111 & -0.36420 & 3.27672 & -1.20531 && -37.745889 & -0.35790 \\
\end{tabular}
\end{ruledtabular}
\end{table*}
\end{turnpage}
\endgroup

\begingroup \squeezetable
\begin{table*}
\caption{Results of the MCHF and CI calculations performed for the C$^-$ $2p^3~^4S^o$ and $^2D^o$ terms. The energies $E$ are in E$_h$, the $S_{sms}$ in $a_0^{-2}$ and the hyperfine parameters in $a_0^{-3}$. The final values are the results of the larger $\lceil 12k\rceil$ calculations on which the impact of the $13^\text{th}$ shell and $l=8$ orbitals have been additively transferred. \label{tab7:resNR_C-}}
\begin{ruledtabular}
\begin{tabular}{clcdddcdddddd}
\multicolumn{2}{c}{model} & \multicolumn{3}{c}{$1s^22s^22p^3~^4S^o$} & \multicolumn{6}{c}{$1s^22s^22p^3~^2D^o$}\\
\cline{1-2}\cline{4-6}\cline{8-13}\\
$nl$ & \multicolumn{1}{c}{$p$} && \multicolumn{1}{c}{$E$} & \multicolumn{1}{c}{$S_{sms}$} & \multicolumn{1}{c}{$a_c$} && \multicolumn{1}{c}{$E$} & \multicolumn{1}{c}{$S_{sms}$} & \multicolumn{1}{c}{$a_l$} & \multicolumn{1}{c}{$a_{sd}$} & \multicolumn{1}{c}{$a_c$} & \multicolumn{1}{c}{$b_q$}\\
\cline{1-2}\cline{4-6}\cline{8-13}\\
&&&\multicolumn{3}{c}{MCHF}&&\multicolumn{6}{c}{MCHF}\\
\\
HF     &            && -37.708844 & -1.60530 & 0.0     && -37.642589 & -1.54597 & 2.35963 & 0.47193 & 0.0     & 0.0 \\
 $ 4 $ &            && -37.862042 & -0.56521 & 0.33257 && -37.810185 & -0.51517 & 2.27217 & 0.50749 & 0.23109 & 0.10808\\
 $ 5 $ &            && -37.876688 & -0.56168 & 0.18050 && -37.827492 & -0.51410 & 2.28209 & 0.54068 & 0.23971 & 0.12912\\
 $ 6 $ &            && -37.884109 & -0.54166 & 0.55357 && -37.836040 & -0.49077 & 2.25909 & 0.52516 & 0.35738 & 0.15402\\
 $ 7 $ &            && -37.887227 & -0.54675 & 0.43564 && -37.839993 & -0.49006 & 2.23484 & 0.52138 & 0.31616 & 0.18659\\
 $ 8k$ &            && -37.888691 & -0.54785 & 0.41389 && -37.841966 & -0.49306 & 2.22493 & 0.52273 & 0.30328 & 0.19303\\
 $ 9k$ &            && -37.889449 & -0.54912 & 0.46136 && -37.842933 & -0.49444 & 2.21796 & 0.52136 & 0.32378 & 0.19962\\
 $10k$ &            && -37.889853 & -0.54975 & 0.45976 && -37.843464 & -0.49534 & 2.21563 & 0.51911 & 0.32252 & 0.20107\\
 $11k$ &            && -37.890085 & -0.55017 & 0.45517 && -37.843751 & -0.49583 & 2.21476 & 0.51910 & 0.32312 & 0.20291\\
 $12k$ &            && -37.890213 & -0.55042 & 0.45714 && -37.843927 & -0.49614 & 2.21462 & 0.51922 & 0.32144 & 0.20238\\
\cline{1-2}\cline{4-6}\cline{8-13}\\
&&&\multicolumn{3}{c}{CI}&&\multicolumn{6}{c}{CI}\\
\\
 $11k$ & $99.0 $    && -37.890143 & -0.54957 & 0.46140 && -37.844634 & -0.48786 & 2.20369 & 0.51222 & 0.30714 & 0.21266\\
       & $99.3 $    && -37.890301 & -0.54774 & 0.45872 && -37.844971 & -0.48375 & 2.20081 & 0.51100 & 0.30452 & 0.21527\\
       & $99.5 $    && -37.890474 & -0.54414 & 0.44761 &&  &  &  &  &  & \\
       & $99.7 $    && -37.890640 & -0.54204 & 0.43942 &&  &  &  &  &  & \\
       & $99.8 $    && -37.890692 & -0.54132 & 0.43823 &&  &  &  &  &  & \\
\\
 $12k$ & $99.0 $    && -37.890271 & -0.54982 & 0.46325 && -37.844826 & -0.48798 & 2.20269 & 0.51165 & 0.30450 & 0.21262\\
       & $99.3 $    && -37.890440 & -0.54797 & 0.46098 && -37.845165 & -0.48386 & 2.19983 & 0.51040 & 0.30176 & 0.21519\\
       & $99.5 $    && -37.890603 & -0.54438 & 0.44919 &&  &  &  &  &  & \\
       & $99.7 $    && -37.890769 & -0.54227 & 0.44080 &&  &  &  &  &  & \\
       & $99.8 $    && -37.890822 & -0.54151 & 0.43945 &&  &  &  &  &  & \\
\\
 $13l$ & $99.0 $    && -37.890429 & -0.55003 & 0.46636 && -37.845003 & -0.48828 & 2.20286 & 0.51161 & 0.30547 & 0.21300\\
\\
\multicolumn{2}{c}{Final} && -37.890980 & -0.54172 & 0.44257 && -37.845343 & -0.48415 & 2.20000 & 0.51036 & 0.30273 & 0.21557 \\
\end{tabular}
\end{ruledtabular}
\end{table*}
\endgroup

\begin{table*}
\caption{Relativistic corrections ($\mu$E$_h$) to the total energies evaluated by BPCI calculations (see text).\label{tab7:bpci}}
\begin{ruledtabular}
\begin{tabular}{clD{.}{.}{2}D{.}{.}{2}D{.}{.}{2}D{.}{.}{2}D{.}{.}{2}D{.}{.}{2}D{.}{.}{2}D{.}{.}{2}}
 \multicolumn{2}{c}{Model} & \multicolumn{5}{c}{C $1s^22s^22p^2$} & \multicolumn{3}{c}{C$^-$ $1s^22s^22p^3$}\\
 \cline{3-7} \cline{8-10}
$nl$ & $p    $ & \multicolumn{1}{c}{$^3P_0$} & \multicolumn{1}{c}{$^3P_1$} & \multicolumn{1}{c}{$^3P_2$} & \multicolumn{1}{c}{$^1D_2$} & \multicolumn{1}{c}{$^1S_0$}
 & \multicolumn{1}{c}{$^4S^o_{3/2}$} & \multicolumn{1}{c}{$^2D^o_{3/2}$} & \multicolumn{1}{c}{$^2D^o_{5/2}$}Ê\\
 \cline{3-7} \cline{8-10}\\
&&\multicolumn{8}{c}{Main spectroscopic terms only (see text)}\\
\\
$10k$ & $99  $ & -14~437.25 & -14~362.42 & -14~239.95 & -14~288.85 & -14~256.62 &  & -14~200.89 & -14~192.92\\
$11k$ & $99  $ & -14~409.01 & -14~334.21 & -14~211.76 & -14~330.16 & -14~296.56\\
$11k$ & $99.5$ & -14~455.75 & -14~380.91 & -14~258.35 & -14~331.96 & -14~299.28\\
 \cline{3-7} \cline{8-10}\\
&&\multicolumn{8}{c}{With additional spectroscopic terms (see text)}\\
\\
$10k$ & $99  $ & -14~437.25 & -14~362.63 & -14~240.02 & -14~288.99 & -14~256.62 & -14~129.53 & -14~201.14 & -14~193.15\\
$11k$ & $99  $ & -14~409.01 & -14~334.42 & -14~211.82 & -14~330.30 & -14~296.56\\
\end{tabular}
\end{ruledtabular}
\end{table*}

\begin{table*}
\caption{Relativistic corrections to the $S_{sms}$ specific mass shift parameter of each considered state evaluated by MCHF-RCI MR-SD calculations. Results are presented in $\mu a_0^{-2}$. \label{tab7:rsms}}
\begin{ruledtabular}
\begin{tabular}{crrrrrrrr}
model& \multicolumn{5}{c}{C~$2p^2$} & \multicolumn{3}{c}{C$^-$~$2p^3$}\\
\cline{2-6}\cline{7-9}
& \multicolumn{1}{c}{$^3P_0$} & \multicolumn{1}{c}{$^3P_1$} & \multicolumn{1}{c}{$^3P_2$} & \multicolumn{1}{c}{$^1D_2$} & \multicolumn{1}{c}{$^1S_0$} & \multicolumn{1}{c}{$^4S^o_{3/2}$} & \multicolumn{1}{c}{$^2D^o_{3/2}$} & \multicolumn{1}{c}{$^2D^o_{5/2}$}\\
\hline
\raisebox{3ex}{}
& \multicolumn{8}{c}{mono-configurational}\\
\raisebox{3ex}{}
RCI  &$  -47 $&$  -47 $&$ -47 $&$  -49 $&$  -53 $&$ -58 $&$ -61 $&$ -61 $\\
DF    &$ -876 $&$ -560 $&$  65 $&$ -243 $&$ -233 $&$ -91 $&$ -57 $&$ -40 $\\
\raisebox{3ex}{}
&  \multicolumn{8}{c}{multi-configurational, mono-reference}\\
\raisebox{3ex}{}
$ 3 $&$   62 $&$  190 $&$  375 $&$  316 $&$  343 $&$  449 $&$  439 $&$  442 $\\
$ 4 $&$ 4555 $&$ 4769 $&$ 5143 $&$ 4933 $&$ 4859 $&$ 5080 $&$ 4993 $&$ 5011 $\\
$ 5 $&$ 3996 $&$ 4228 $&$ 4636 $&$ 4388 $&$ 4413 $&$ 4595 $&$ 4678 $&$ 4700 $\\
$ 6 $&$ 3837 $&$ 4067 $&$ 4473 $&$ 4259 $&$ 4376 $&$ 4230 $&$ 4131 $&$ 4154 $\\
$ 7 $&$ 3918 $&$ 4157 $&$ 4579 $&$ 4338 $&$ 4324 $&$ 4305 $&$ 4180 $&$ 4210 $\\
$ 8 $&$ 3939 $&$ 4175 $&$ 4593 $&$ 4337 $&$ 4338 $&$ 4329 $&$ 4272 $&$ 4304 $\\
\raisebox{3ex}{}
&  \multicolumn{8}{c}{multi-configurational, multi-reference}  \\
\raisebox{3ex}{}
$ 3 $&$   98 $&$  225 $&$  412 $&$  351 $&$  390 $&$  441 $&$  442 $&$  449 $\\
$ 4 $&$ 4750 $&$ 4967 $&$ 5349 $&$ 5128 $&$ 5134 $&$ 5201 $&$ 5106 $&$ 5128 $\\
$ 5 $&$ 4167 $&$ 4402 $&$ 4817 $&$ 4559 $&$ 4633 $&$ 4713 $&$ 4794 $&$ 4820 $\\
$ 6 $&$ 3999 $&$ 4233 $&$ 4647 $&$ 4426 $&$ 4570 $&$ 4347 $&$ 4241 $&$ 4268 $\\
$ 7 $&$ 4081 $&$ 4323 $&$ 4752 $&$ 4506 $&$ 4509 $&$  $\\
\raisebox{3ex}{}
final&$ 4102 $&$ 4341 $&$ 4766 $&$ 4506 $&$ 4523 $&$ 4446 $&$ 4382 $&$ 4418 $
\end{tabular}
\end{ruledtabular}
\end{table*}

\subsection{Breit-Pauli calculations}\label{sec7:bpci}

A first way to include relativistic effects is to use the Breit-Pauli Hamiltonian that includes the $1/c^{2}$ relativistic correction operators to the non-relativistic atomic Hamiltonian~\cite{Froetal:97a}.

Since the radiative transitions we consider are essentially authorized by $L$ and $S$ mixing, we need to have a good description of the term mixing. On the other hand, it is the calculation of the scalar relativistic effects that is needed for estimating the relativistic effects on the electron affinity since the fine structures of the involved species are usually known experimentally. We therefore choose two distinct Breit-Pauli models. The first BPCI CSF lists are used for the term separation and detachment thresholds corrections while the second approach is used for the transition probabilities calculations.

Focusing on the correlation, we merge the MR$_{99}$-I$\lceil 10k\rceil$ lists of the studied terms for both C$^-$ and C. Then we extend this model by adding the CSFs interacting to first order with the CSFs $2p3p$ $LS$, \mbox{$LS=\ ^3D,\ ^3S,\ ^1P$} for C and the CSFs $2p^{2}\ 3p$ $LS$, $LS=\ ^2P^o,\ ^2F^o,\ ^4D^o,\ ^4P^o$ for C$^-$. For the neutral carbon we test the impact of additional correlation on the relativistic corrections by using the \mbox{MR$_{99}$-I$\lceil 11k\rceil$} and \mbox{MR$_{99.5}$-I$\lceil 11k\rceil$} spaces. We finally diagonalize the Breit-Pauli Hamiltonian in those CSF spaces using the corresponding active sets optimized in the non-relativistic calculations.
The relativistic corrections to the energy are summarized in Table~\ref{tab7:bpci}.
We see that the effect of the additional $LS$ mixing on the energy levels is so small that only the corrections on the fine structures are meaningful.

For the reasons expressed above, we also perform BPCI calculations that
focus on term mixing.
For each $LSJ$ and active set $\lceil nl\rceil$ ($nl=4f-12k$ for C and $4f-8k$ for C$^-$),
the MR$_{98}$-I list is merged with the MR-I set obtained using the reference containing all allowed $LS$ couplings of the $2s\rightarrow 3d$, $2p\rightarrow 3p$ and $2s^2\rightarrow 2p^2$ excitations from the main configuration.

\subsection{Relativistic configuration interaction calculations}\label{sec7:rci}

\begin{table*}
\caption{Relativistic corrections on $A\frac{I}{\mu_I}$ [kHz$/\mu_N$] and $B/Q$ [kHz$/$barn]  of each considered state evaluated by MCHF-RCI MR-SD calculations. \label{tab7:rhfs}}
\begin{ruledtabular}
\begin{tabular}{crrrrrrrrrrrr}
model& \multicolumn{4}{c}{$2p^2~^3P$} & \multicolumn{2}{c}{$2p^2~^1D$} & \multicolumn{2}{c}{$2p^3~^4S^o$} & \multicolumn{4}{c}{$2p^3~^2D^o$}\\
\cline{2-5}\cline{6-7}\cline{8-9}\cline{10-13}
($n_{max}$) & \multicolumn{1}{c}{$A_1$} & \multicolumn{1}{c}{$B_1$} & \multicolumn{1}{c}{$A_2$} & \multicolumn{1}{c}{$B_2$} & \multicolumn{1}{c}{$A_2$} & \multicolumn{1}{c}{$B_2$}  & \multicolumn{1}{c}{$A_{3/2}$} & \multicolumn{1}{c}{$B_{3/2}$}  & \multicolumn{1}{c}{$A_{3/2}$} & \multicolumn{1}{c}{$B_{3/2}$}  & \multicolumn{1}{c}{$A_{5/2}$} & \multicolumn{1}{c}{$B_{5/2}$} \\
\hline
\raisebox{3ex}{}
& \multicolumn{12}{c}{mono-configurational}\\
\raisebox{3ex}{}
RCI &$  93 $&$  2 $&$ 118 $&$  90 $&$-141 $&$ -82 $&$    2 $&$  0.1 $&$   69 $&$ 855 $&$  -24 $&$ 0 $\\
DF   &$-158 $&$-10 $&$ 164 $&$-177 $&$ -26 $&$ 150 $&$ -122 $&$ -0.2 $&$ -464 $&$ 550 $&$  371 $&$ 0 $\\
\raisebox{3ex}{}
& \multicolumn{12}{c}{multi-configurational, mono-reference}\\
\raisebox{3ex}{}
$3  $&$ -221 $&$  -51 $&$ -52 $&$   38 $&$ -166 $&$-130 $&$ -404 $ &$ -0.4 $ &$ -62 $&$ 1317 $&$-113 $&$ 26 $\\
$4  $&$ -226 $&$  -31 $&$ 157 $&$ -111 $&$  -23 $&$  61 $&$ -301 $ &$ -0.1 $ &$ -18 $&$ 1276 $&$  50 $&$ 24 $\\
$5  $&$ -256 $&$  -34 $&$ 152 $&$ -121 $&$  -13 $&$  70 $&$ -322 $ &$  0.0 $ &$ -16 $&$ 1280 $&$  68 $&$ 23 $\\
$6  $&$ -254 $&$  -35 $&$ 163 $&$ -123 $&$   -3 $&$  64 $&$ -319 $ &$  0.2 $ &$ -19 $&$ 1356 $&$  73 $&$ 27 $\\
$7  $&$ -279 $&$  -39 $&$ 158 $&$ -135 $&$    6 $&$  71 $&$ -357 $ &$ -0.1 $ &$ -19 $&$ 1393 $&$  77 $&$ 36 $\\
$8  $&$ -285 $&$  -35 $&$ 160 $&$ -146 $&$   11 $&$  85 $&$ -368 $ &$ -0.1 $ &$ -16 $&$ 1392 $&$  81 $&$ 36 $\\
\raisebox{3ex}{}
& \multicolumn{12}{c}{multi-configurational, multi-reference} \\
\raisebox{3ex}{}
$3  $&$ -221 $&$  -52 $&$ -52 $&$   38 $&$ -168 $&$-132 $&$ -414 $ &$ -0.4 $ &$ -66 $&$ 1267 $&$ -117 $&$ 25 $\\
$4  $&$ -217 $&$  -34 $&$ 166 $&$ -106 $&$  -24 $&$  52 $&$ -289 $ &$ -0.1 $ &$ -26 $&$ 1210 $&$   52 $&$ 25 $\\
$5  $&$ -246 $&$  -37 $&$ 162 $&$ -115 $&$  -14 $&$  59 $&$ -306 $ &$  0.0 $ &$ -25 $&$ 1218 $&$   71 $&$ 24 $\\
$6  $&$ -242 $&$  -38 $&$ 174 $&$ -116 $&$   -4 $&$  52 $&$ -298 $ &$  0.1 $ &$ -30 $&$ 1291 $&$   78 $&$ 28 $\\
$7  $&$ -266 $&$  -42 $&$ 170 $&$ -128 $&$    6 $&$  57 $\\
\raisebox{3ex}{}
final&$ -272 $&$  -38 $&$ 172 $&$ -138 $&$   10 $&$  71 $&$ -347 $ &$ -0.1 $ &$ -27 $&$ 1328 $&$   86 $&$ 38 $
\end{tabular}
\end{ruledtabular}
\end{table*}

We use essentially the same method as in~\cite{Jonetal:10a}. First, we perform reference MCHF calculations with all single and double configuration excitations (SD) of the ground state in active sets ranging from $\lceil 3d \rceil$ to $\lceil 8k \rceil$. The resulting non-relativistic radial orbitals $P_{nl}(r)$ are then converted to Dirac spinors using the Pauli approximation
\beqa
P_{n\kappa}(r) &=& P_{nl}(r)\\
Q_{n\kappa}(r) &=& \frac{\alpha}{2} \left( \frac{d}{dr} + \frac{\kappa}{r} \right) P_{nl}(r)
\eeqa
where $\alpha$ is the fine structure constant and $\kappa$ is defined
\beq
\kappa = \left\{\begin{array}{cl} -l-1 &\qquad \textrm{when} \qquad j=l+1/2\\
l &\qquad \textrm{when} \qquad j=l-1/2 \end{array}\right.
\eeq
We finally perform the corresponding RCI calculations using the set of SD excitations of the main configuration.
Larger configuration sets are explored by means of non-relativistic CI and RCI calculations using the references
\beqa
\text{MR$($C}^-)&=&\{1s^22s^22p^3, 1s^22s^12p^33d^1\}\, ,\\
\text{MR$($C})\phantom{{}^-}&=&\{1s^22s^22p^2, 1s^22s^12p^23d^1, 1s^22p^4\}.
\eeqa
The relativistic effects are estimated from the differences between the non-relativistic CI and corresponding RCI results.

In Table~\ref{tab7:rsms}, we compare the relativistic corrections on $S_{sms}$ obtained with our calculations to the ones deduced from single configuration calculations (RCI--HF and DF--HF).
It seems that correlation plays an important role in the estimation of these corrections, as could be expected from an operator that measures the correlation between the momenta of the electrons. The convergence of the mono-reference approach with $n$ is sufficient. However, we see a large change between the mono- and multi-reference approaches.
In Table~\ref{tab7:rhfs}, we present the corrections for $A\frac{I}{\mu_I}$ and $B/Q$ that are both independent of the nuclear spin $I$ and multipole moments~($\mu_I, Q$).

Similarly to the non-relativistic calculations, we note that for neutral carbon, the impact of the $7^\text{th}$ and $8^\text{th}$ shells is not much affected by the choice of reference. We then estimate the final value as in the non-relativistic case.

\begin{table}
\caption{$A\frac{I}{\mu_I}$ [kHz$/\mu_N$] and $B/Q$ [kHz$/$barn] theoretical values for carbon $^3P$, $^1D$ and C$^-$ $^4S^o$, $^2D^o$.\label{tab7:indepHFS}}
\begin{ruledtabular}
\begin{tabular}{lrrclrr}
\multicolumn{3}{c}{C}&&\multicolumn{3}{c}{C$^-$}\\
\cline{1-3}\cline{5-7}
\multicolumn{1}{c}{state} & \multicolumn{1}{c}{$A\frac{I}{\mu_I}$} & \multicolumn{1}{c}{$B/Q$} && \multicolumn{1}{c}{state} & \multicolumn{1}{c}{$A\frac{I}{\mu_I}$} & \multicolumn{1}{c}{$B/Q$}\\
\hline
\multicolumn{7}{c}{Non-relativistic}\\
\\
$2p^2~^3P_1 $&$   2~296 $&$  74~184 $&&$ 2p^3~^4S^o_{3/2} $&$   9~394 $&$      0 $\\
$2p^2~^3P_2 $&$ 105~883 $&$-148~367 $&&$ 2p^3~^2D^o_{3/2} $&$  53~836 $&$-35~455 $\\
$2p^2~^1D_2 $&$ 156~317 $&$ 283~206 $&&$ 2p^3~^2D^o_{5/2} $&$ 107~317 $&$-50~650 $\\
\hline
\multicolumn{7}{c}{+ relativistic corrections}\\
\\
$2p^2~^3P_1 $&$   2~024 $&$  74~145 $&&$ 2p^3~^4S^o_{3/2} $&$   9~048 $&$      0 $\\
$2p^2~^3P_2 $&$ 106~055 $&$-148~505 $&&$ 2p^3~^2D^o_{3/2} $&$  53~809 $&$-34~128 $\\
$2p^2~^1D_2 $&$ 156~327 $&$ 283~276 $&&$ 2p^3~^2D^o_{5/2} $&$ 107~403 $&$-50~613 $
\end{tabular}
\end{ruledtabular}
\end{table}

\section{Results and comparison to experiment}\label{sec7:expth}

\subsection{Hyperfine Structures}\label{sec7:expthHFS}

\begin{table*}
\caption{Comparison of our calculated hyperfine constants of $^{13}$C to other works. The experimental values are adjusted according to our analysis of the off-diagonal $JJ'$ interaction. All values are in MHz.\label{tab7:revHFSexp13}}
\begin{ruledtabular}
\begin{tabular}{llllclll}
&\multicolumn{3}{c}{$^{13}$C} &&\multicolumn{3}{c}{$^{13}$C$^-$}\\
\cline{2-4}\cline{6-8}
& \multicolumn{1}{c}{$A_{1}(^3P)$} & \multicolumn{1}{c}{$A_2(^3P)$} & \multicolumn{1}{c}{$A_2(^1D)$}&& \multicolumn{1}{c}{$A_{3/2}(^4S^o)$} & \multicolumn{1}{c}{$A_{3/2}(^2D^o)$} & \multicolumn{1}{c}{$A_{5/2}(^2D^o)$}\\
\hline
Original exp.\footnote{Reference \cite{Woletal:70a}.}     &$ 2.838(17) $&$ 149.055(10) $&&\\
This work   &$ 2.84    $&$ 148.99   $&$ 219.61 $ &&$ 12.71 $&$ 75.59 $&$ 150.88 $ \\
Prev. work\footnote{Reference \cite{Jonetal:96a}.} &$ 2.28 $&$ 148.1 $& &
\end{tabular}
\end{ruledtabular}
\end{table*}

In this work, we focus on the isotopes $13$ and $11$ of carbon, respectively of nucleus spin $1/2$ and $3/2$. The $^{11}$C nucleus decay into $^{11}B$ by $e^+$-emission with a half-lifetime of $20.4$~minutes.
Haberstroh \etal \cite{Habetal:64a} and Wolber \etal \cite{Woletal:70a} performed experimental studies of the hyperfine structures of the carbon ground state of $^{11}$C and $^{13}$C, respectively. In the latter article, a magnetic dipole-moment of $^{11}$C of $-0.964(1)~\mu_N$ was deduced from the then available $\mu$($^{13}$C) value. We update this estimation by using the modern $\mu$($^{13}$C) value~\cite{Sto:05a} combined with the two measured $A(^3P_2)$ constants:
\beq
\mu(^{11}\text{C}) = \left(\frac{A_J(^{11}\text{C})\ I_{11}\ \mu(^{13}\text{C})}{A_J(^{13}\text{C})\ I_{13}}\right)_{exp} = -0.9642(2)\ \mu_N\, .
\eeq
The error on this value is now dominated by the accuracy of the $A(^3P_2)$ hyperfine constants measurements. 

As mentioned in Section~\ref{sec7:nrcalc}, it is difficult to have a rigorous estimation of the uncertainty on the hyperfine parameters.
We however advance a learned guess of their reliability. First, we see in Table~\ref{tab7:resNR_C} that the integrals $a_l$, $a_{sd}$ and $b_q$ of C change less than $0.05\%$ after the addition of the $\lceil 13l\rceil$ correction. The $a_c$ parameter of the $^3P$ is only slightly more affected ($\sim 0.1\%$). These effects are representative of the accuracy of our results for neutral carbon.
In the case of C$^-$ we face two additional limitations: the structure of the negative ion converges more slowly and we are limited in our expansions. Moreover, only the most troublesome contact term is responsible for the non-relativistic HFS of $2p^3~^4S^o$. For these reasons, and comparing the values of Table~\ref{tab7:resNR_C-} with results obtained with the active set $\lceil 10k\rceil$, we must allow for relative uncertainties on the HFS parameters roughly ten times larger for C$^-$ than for C.

In Table~\ref{tab7:indepHFS}, we present the non-relativistic $A\frac{I}{\mu_I}$ and $B/Q$ results calculated using the final values of $a_l, a_{sd}, a_c$ and $b_q$ of Tables~\ref{tab7:resNR_C} and~\ref{tab7:resNR_C-}. In the same table, we add the relativistic corrections of Table~\ref{tab7:rhfs} to those values. The $A(^3P_1)$ constant is the place of severe compensations between the orbital ($a_l$) and spin-dipole ($a_{sd}$) contributions so that  the uncertainties on those sum up to an error of the order of $10^2$~kHz$/\mu_N$. The other nuclear-parameters-independent hyperfine constants of neutral carbon suffer of a non-relativistic uncertainty of about \mbox{$10-10^2$~kHz$/\mu_N$}. These are larger than the fluctuations observed in Table~\ref{tab7:rhfs}. As far as C$^-$ is concerned, on the one hand the $^4S^o$ hyperfine structure is essentially due to the contact term, itself arising only from correlation effects, and on the other hand, the $^2D^o$ hyperfine constants are small but the achieved convergence of the calculations is less good. Therefore the relative non-relativistic uncertainties on C$^-$ hyperfine structures are larger as they sum up to about $50-100~$kHz$/\mu_N$~(kHz$/$barn). We conclude that the reliability of all normalized hyperfine constants is of the order of $10^2$~kHz$/\mu_N$ (kHz$/$barn) with the exception of the $B(^4S^o)$ that is certainly negligible.


\begin{table*}
\caption{Comparison of our calculated hyperfine constants of $^{11}$C to other works. The experimental values are adjusted according to our analysis of the off-diagonal $JJ'$ interaction and of $B(^3P_1)/B(^3P_2)$. All values are in MHz.\label{tab7:revHFSexp11}}
\begin{ruledtabular}
\begin{tabular}{lllllll}
&\multicolumn{6}{c}{$^{11}$C} \\
\cline{2-7}
& \multicolumn{1}{c}{$A_1(^3P)$} & \multicolumn{1}{c}{$B_1(^3P)$} & \multicolumn{1}{c}{$A_2(^3P)$} & \multicolumn{1}{c}{$B_2(^3P)$} & \multicolumn{1}{c}{$A_2(^1D)$} & \multicolumn{1}{c}{$B_2(^1D)$}\\
\hline
Original exp.\footnote{Reference \cite{Habetal:64a}.}     &$ -1.308(24) $&$ 2.475(14) $&$ -68.203(7) $&$ -4.949(28)$\\
This work &$ -1.30 $&$ 2.474 $&$ -68.17     $&$ -4.955 $&$ -100.49 $&$ 9.450 $\\
\hline
&\multicolumn{6}{c}{$^{11}$C$^-$} \\
\cline{2-7}
& \multicolumn{1}{c}{$A_{3/2}(^4S^o)$} & \multicolumn{1}{c}{$B_{3/2}(^4S^o)$} & \multicolumn{1}{c}{$A_{3/2}(^2D^o)$} & \multicolumn{1}{c}{$B_{3/2}(^2D^o)$} & \multicolumn{1}{c}{$A_{5/2}(^2D^o)$} & \multicolumn{1}{c}{$B_{5/2}(^2D^o)$}\\
\hline
This work &$ 5.82 $&$ \approx 0 $&$ 34.59 $&$ -1.139 $&$ 69.04 $&$ -1.688 $\\
\end{tabular}
\end{ruledtabular}
\end{table*}

Our results are compared with observations in Tables~\ref{tab7:revHFSexp13} and~\ref{tab7:revHFSexp11} for C and C$^-$ respectively. We observe a good agreement with experiment, better than expected from the above discussion. This represents a significant improvement compared to the theoretical study of J\"onsson~\etal \cite{Jonetal:96a}. 

The observed hyperfine splittings arise from the diagonal hyperfine interaction, parametrized by the $A_J$ and $B_J$ constants, and, to higher order, from the non-diagonal ($JJ'$) interaction of states of same $F$. If only two levels are involved, one must diagonalize the matrix
\beq
\left(\begin{array}{cc} 0 & W(JJ'; F) \\ W(JJ'; F) & \Delta_{JJ'}E(LS\ F) \end{array} \right)
\eeq
where $\Delta_{JJ'}E(LS\ F)=E(LSJ'F)-E(LSJF)$ is dominated by the fine structure splitting and $W(JJ'; F)$ is governed by the off-diagonal hyperfine constants -- here $A_{J,J-1}$, $B_{J,J-1}$ and $B_{J,J-2}$~(see Section~\ref{sec3:HFS}). The off-diagonal electric quadrupole interaction is negligible and, at the non-relativistic level, we obtain for C($^3P$)
\beqa
I A_{1,0}/\mu_I &=& 50.47~\text{MHz}/\mu_N\, , \\
I A_{2,1}/\mu_I &=& 62.71~\text{MHz}/\mu_N\, , \\
\eeqa
while for C$^-$($^2D^o$) we have
\beq
I A_{5/2,3/2}/\mu_I = 34.20~\text{MHz}/\mu_N.
\eeq
The hyperfine interaction between states belonging to different terms is negligible.

Wolber \etal \cite{Woletal:70a} measured two hyperfine splittings in the $^{13}$C $^{3}P_J$ multiplet, allowing the determination of the $A_1$ and $A_2$ diagonal constants but not of the off-diagonal constants so that they had to deduce the contribution of the $JJ'$--interaction theoretically. The level shifts that they obtained from their computations are significantly higher than ours. However, the $A_J$ constants that reproduce the experimental hyperfine splittings when using our results for the $JJ'$-interaction, $A_1=2.829(17)$ MHz and $A_2=149.052(10)$ MHz, do not differ largely from the experimental constants presented in Table~\ref{tab7:revHFSexp13}.

Haberstroh~\etal \cite{Habetal:64a} measured three hyperfine splittings for $^{11}$C $^3P_J$, which is insufficient for determining all four $A_J$ and $B_J$ of this term. Hence they deduced the value of $B_1$ from the relation
$B_2/B_1=-2$
which is only valid in the Hartree-Fock model. From Table~\ref{tab7:indepHFS}, we see that this formula holds very well at the non-relativistic level
but that, including relativistic corrections, we have
\beq
B_2/B_1=-2.0029\, .\label{eq:B1B2rat}
\eeq
The effect of the refined $B_2/B_1$ ratio and $JJ'$-interactions cancel each other in the estimation of the diagonal hyperfine constants so that the resulting $A_J$ constants do no differ significantly from the experimental ones quoted in Table~\ref{tab7:revHFSexp11}. For the electric quadrupole interaction, the accuracy of our results is such that we can safely update the electric quadrupole moment of the $^{11}$C nucleus with the formula
\beq
Q(^{11}\text{C}) =\frac{\left( B_2(^{11}\text{C}) \right)_{exp}}{\left( B_2/Q \right)_{th}}.
\eeq
Using the $B_2$ constant of Haberstroh~\etal \cite{Habetal:64a}, we obtain a value of $+0.03333(19)_{exp}(2)_{th}$~barns but if we use our theoretical parameters in the analysis of the observations, we obtain
\beq
Q(^{11}\text{C}) =  +0.03336(19)_{exp}(2)_{th} \text{ barns}\, .\label{eq:QC11}
\eeq
This value is used for estimating the theoretical $B_J$ constants of this work presented in Table~\ref{tab7:revHFSexp11}. The difference between theory and experiment for the $B_2$ constant follows directly from the fact that (\ref{eq:QC11}) includes the refinements of the theoretical parameters needed in the analysis of the observed hyperfine splittings.

Let us mention the previous calculations of the $b_q$ parameter (we get $b_q=0.6314~a_0^{-3}$):
$b_q=0.6325~a_0^{-3}$~\cite{SunOls:92a}, $b_q=0.6319~a_0^{-3}$~\cite{Jonetal:93a}.
Using the experimental constant $B_2$ quoted in Table~\ref{tab7:revHFSexp11}, Sundholm and Olsen~\cite{SunOls:92a} proposed $Q(^{11}\text{C})=+0.03327(24)$~barns which would only tenuously agree with our estimation if the $(B_2(^{11}\text{C}))_{exp}$ value was to be improved.

In the case of C$^-$, the small $^2D^o$ fine structure (1.75~cm$^{-1}$, see below), leads to  $JJ'$-interaction shifts on the energy levels that are roughly 10 times larger than in the neutral atom ground term, \ie of the order of 0.1~MHz.

\subsection{Energy differences}\label{sec7:endiff}

Table~\ref{tab7:endiff} presents several calculated energy separations and compares them to other works. 

Our C$^-$ term splitting is in very good agreement with experiment but, as will be seen below, this is partially accidental.

Our results on the neutral atom ground configuration level spacings are systematically better than the ones of Froese-Fischer and Tachiev~\cite{FroTac:04a}. It indicates that, in this context, our relativistic corrections are reliable.
For the $^{3}P$ fine structure, we obtain as accurate results as recent fully relativistic calculations~\cite{Kozetal:09a}.

Our systematic procedure is not particularly efficient for predicting the negative ion binding energy. In particular, for the $^4S^o$ detachment threshold, the coupled-cluster approaches are much more impressive~\cite{Olietal:99a,Kloetal:10a}. The recent value of Klopper \etal \cite{Kloetal:10a} indeed achieves a sub-meV ($<8$ cm$^{-1}$) agreement with the experimental electron affinities for all first- and second-period atoms (H-Ne). A similar accuracy had already been achieved more than 10 years before by de Oliveira \etal \cite{Olietal:99a} for the second and third period $p-$block atoms. 

By trying various extrapolation schemes on our C$^-$ calculations,
we explain up to $\sim 20$~cm$^{-1}$ of the difference between our calculation of the $^4S^o$ binding energy and the experimental value (about $5$ cm$^{-1}$ for each $n$ and $l$ extrapolations, and about another $10$~cm$^{-1}$ for the extrapolation to a complete active set). Turning to the relativistic effects calculations, we see that the scalar contributions calculated with the CC methods give $-21.54$~cm$^{-1}$~\cite{Kloetal:10a} and  $-22.83$ cm$^{-1}$~\cite{Olietal:99a} while we obtain $-37.95$~cm$^{-1}$. The extrapolation being reliable to about a couple of tenth of percents and since the additional expected contributions are of the order of the cm$^{-1}$, we conclude that our BPCI relativistic corrections are still unbalanced. The problem of our relativistic corrections on the detachment thresholds is confirmed by the fact that, looking to Table~\ref{tab7:bpci}, they are not well converged.

\begin{table*}
\caption{Comparison of the theoretical and experimental energy level separations. The MCHF -- CI calculations of the energy differences involving C$^-$ $^4S^o$ and $^2D^o$ are obtained with $p=99.8\%$ and $99.3\%$, respectively, while we take $p=99.95\%$ for neutral carbon transitions energies. Relativistic corrections ($+$rel) are estimated from Table~\ref{tab7:bpci} and the $+\lceil 13l\rceil$ column corresponds to the final results of Tables~\ref{tab7:resNR_C} and~\ref{tab7:resNR_C-}.  All values are given in cm$^{-1}$.\label{tab7:endiff}}
\begin{ruledtabular}
\begin{tabular}{cccD{.}{.}{3}D{.}{.}{3}D{.}{.}{3}D{.}{.}{3}D{.}{.}{6}D{.}{.}{3}D{.}{.}{3}}
\multicolumn{3}{c}{State} & \multicolumn{4}{c}{This work} & \multicolumn{1}{c}{Exp.} & \multicolumn{2}{c}{Prev. Th.}\\
\cline{4-7}\cline{9-10}
& &&\multicolumn{1}{c}{HF} & \multicolumn{1}{c}{MCHF -- CI} & \multicolumn{1}{c}{$+$rel} & \multicolumn{1}{c}{+$\lceil 13l \rceil$} \\
\hline
\multicolumn{3}{c}{C$^-(^4S^o_{3/2})$} & 0.0 & 0.0 & 0.0 & 0.0 & 0.0\\
\multicolumn{3}{c}{$\begin{array}{c}
\text{C}^-(^2D^o_{3/2}) \\
\text{C}^-(^2D^o_{5/2})
\end{array} $} & 14~541.11 & 9~936.73 &
\begin{array}{r}
9~921$.$01 \hspace*{-0.6cm}\\
9~922$.$76 \hspace*{-0.6cm} \end{array} &
\begin{array}{r}
9~916$.$63 \hspace*{-0.6cm}\\
9~918$.$39 \hspace*{-0.6cm} \end{array} &  9~913.5(82)\footnote{Reference \cite{Fel:77a}.}  \\
\\
\multicolumn{3}{c}{C$(^3P_0)$}     &      0.0  &      0.0  &      0.0  &  0.0     & 0.0 &      0.0\\
\multicolumn{3}{c}{C$(^3P_1)$}     &      0.0  &      0.0  &     16.39 &     16.39  &     16.42\footnote{Reference \cite{Kleetal:98a}.} &     16.33^\text{c} & 16.4^\text{d}\\
\multicolumn{3}{c}{C$(^3P_2)$}     &      0.0  &      0.0  &     43.31 &     43.31  &     43.41^\text{b} &     43.03\footnote{Reference \cite{FroTac:04a}.} & 43.3\footnote{Reference \cite{Kozetal:09a}.}\\
\multicolumn{3}{c}{C$(^1D_2)$}     & 12~573.20 & 10~193.42 & 10~220.55 & 10~213.76  & 10~192.63\footnote{Reference \cite{Moo:93a}.} & 10~268.23^\text{c} \\
\multicolumn{3}{c}{C$(^1S_0)$}     & 30~508.75 & 21~657.45 & 21~691.79 & 21~682.40  & 21~648.01^\text{e} & 21~818.60^\text{c} \\
\hline
C$^- $&$-$&$$C& \multicolumn{7}{c}{Detachment thresholds} \\
\\
$^4S^o_{3/2} $&$-$&$^3P_0$ &  4~438.80 & 10~200.50 & 10~132.96 & 10~141.49 & 10~179.68(16)\footnote{Reference \cite{Schetal:98a}.} & 10~184.61\footnote{Reference \cite{Kloetal:10a}.} & 10~185.8\footnote{Reference \cite{Olietal:99a}.}\\
$^4S^o_{3/2} $&$-$&$^1D_0$ & 17~012.00 & 20~391.04 & 20~356.04 & 20~357.77 & 20~372.31(16)^\text{f} &  & \\
$^4S^o_{3/2} $&$-$&$^1S_0$ & 34~947.55 & 31~853.18 & 31~825.28 & 31~824.42 & 31~827.69(16)^\text{f} &  & \\
$^2D^o_{3/2} $&$-$&$^3P_0$ &-10~102.31 &    240.39 &    194.77 &    207.67 & 266.2(81)  & \multicolumn{1}{l}{$\phantom{00~}436\footnote{Reference \cite{Zhoetal:05a}.}$} \\
$^2D^o_{3/2} $&$-$&$^1D_2$ &  2~470.89 & 10~440.27 & 10~420.99 & 10~427.10 & 10~458.8(81)^\text{a} & \\
$^2D^o_{3/2} $&$-$&$^1S_0$ & 20~406.44 & 21~898.72 & 21~886.54 & 21~890.05 & 21~914.2(81)^\text{a} &
\end{tabular}
\end{ruledtabular}
\end{table*}

Aside a possible unbalance in the relativistic effects estimation, our error is roughly proportional to the correlation contribution. We see that the differences between the HF and experimental energy separations (see Table~\ref{tab7:endiff}) are reproduced to $\sim 0.1-0.7\%$, which is about the percentage of the C$(^3P)$ correlation energy we get. It means that $p$ and $n$ are good indicators of the percentage of the correlation effects included in a model. However, the uncertainty on our relativistic corrections and on the $^2D^o$ missing correlation is too large for an extrapolation based on this observation to be useful, \eg for improving the experimental determination of the position of the $^2D^o_J$ levels. Indeed, a $0.5\%$ uncertainty on our calculated correlation energies, which is no overestimation, reflects in corrections ranging from $\sim 60$ cm$^{-1}$ in the case of the largest HF-experiment discrepancy, to about $7.5$ cm$^{-1}$ for the $^2D^o - {^1S}$ threshold, \ie of the same order of magnitude than the experimental uncertainty.

We would like to stress another advantage of using the number of correlation layers $n$ and $p$ as parameters for preserving the balance between systems having different numbers of electrons. Observing that our models converge toward the exact solution of the Schr\"odinger equation,
 a larger number of electrons demands larger active sets, and
that for a given orbital set, $p$ is roughly proportional to the amount of correlation in the model,
we affirm that the results obtained with increasing $n$ and $p$ will most often underestimate the photodetachment thresholds. In other words, the detachment thresholds are valuable references for estimating the accuracy of the calculations since their behavior is monotone (as the level energies themselves). 

\begin{table*}
\caption{Comparison of our isotope shifts ($A'=13, A=12$), in m$^{-1}$, on various positive energy separations. The MCHF -- CI calculations of the energy differences involving C$^-$ $^4S^o$ and $^2D^o$ are obtained with $p=99.8\%$ and $99.3\%$, respectively, while $p=99.95\%$ for neutral carbon transitions. Relativistic corrections ($+$rel) are estimated from Table~\ref{tab7:rsms} and the $+\lceil 13l\rceil$ column corresponds to the final results of Tables~\ref{tab7:resNR_C} and~\ref{tab7:resNR_C-}. \label{tab7:IS}}
\begin{ruledtabular}
\begin{tabular}{D{-}{-}{-1}ddddddD{.}{.}{8}}
&\multicolumn{3}{c}{SMS} & \multicolumn{1}{c}{NMS} & \multicolumn{3}{c}{IS} \\
\cline{2-4}\cline{6-8}
\multicolumn{1}{c}{trans.} & \multicolumn{1}{c}{MCHF -- CI} & \multicolumn{1}{c}{$+$rel} & \multicolumn{1}{c}{$+\lceil 13l \rceil$} & \multicolumn{1}{c}{Exp.} & \multicolumn{1}{c}{This work} & \multicolumn{1}{c}{Other th.} & \multicolumn{1}{c}{Exp}\\
\hline
\raisebox{4ex}{}%
&\multicolumn{7}{c}{$IS$ on the C$^-$ terms separations}\\
\raisebox{4ex}{}%
{^4S}^o_{3/2} \ -\ ^2D^o_{3/2} &  -4.965 &  -4.960 &  -4.953 & 3.500 & -1.454 \\
\raisebox{4ex}{}%
&\multicolumn{7}{c}{$IS$ on the C$^-$ detachment thresholds}\\
\raisebox{4ex}{}%
{^4S}^o_{3/2} \ -\ ^3P_0 & -10.656 & -10.629 & -10.630 & 3.592 & -7.038 & -8.7\footnote{Non-relativistic coupled-cluster calculations~\cite{Kloetal:10a}.} \\
^2D^o_{3/2} \ -\ ^1D_2 &  -8.865 &  -8.875 &  -8.878 & 3.690 & -5.185 \\
\raisebox{4ex}{}%
&\multicolumn{7}{c}{$IS$ on the C terms separations}\\
\raisebox{4ex}{}%
{^3P_0} \ -\ ^1D_2 &  -3.021 & -3.052 & -3.047 & 3.597 & +0.550 & +0.505\footnote{Calculations using the Dirac-Breit Hamiltonian and the relativistic mass shift operator~\cite{Kozetal:09a}. Note that they used a different sign convention than ours (Kozlov, private communication).}  \\
^3P_0 \ -\ ^1S_0 &  -3.507 & -3.540 & -3.536 & 7.640 & +4.103 & +4.672^\text{b} &  \\
&&&&&& +4.374\footnote{MCHF -- CI calculations~\cite{Caretal:95a}. SMS $ = -3.266$ m$^{-1}$.} \\
\raisebox{4ex}{}%
&\multicolumn{7}{c}{$IS$ on the C$(^3P)$ fine structure}\\
\raisebox{4ex}{}%
{^3P_1} \ -\ ^3P_2 &   0     & -0.033 & -0.033 & 0.010 & -0.023 & +0.020^\text{b} & +0.0137(10)\footnote{Reference \cite{Kleetal:98a}.} \\
&&&&& +0.014\footnote{Our results combined with the relativistic corrections of Veseth~\cite{Ves:85a} (see text).} & +0.015\footnote{MBPT calculations of the relativistically corrected mass shift operator~\cite{Ves:85a}. Note that those values are quoted with the wrong sign in reference~\cite{Cooetal:86a}, as pointed out in reference~\cite{Ves:87a}.} & +0.0180(43)\footnote{Reference \cite{Cooetal:86a}.} \\ 
{^3P}_0 \ -\ ^3P_1 &   0     & -0.019 & -0.019 & 0.006 & -0.013 & +0.009^\text{b} & +0.0077(7)\footnote{Measurements of~\cite{YamSai:91a} and hyperfine splittings of~\cite{Woletal:70a}.} \\
&&&&&& +0.010^\text{f} & +0.0057(83)^\text{g}.
\end{tabular}
\end{ruledtabular}
\end{table*}

\subsection{Mass isotope shifts}\label{sec7:ms}

Table~\ref{tab7:IS} reports the results for the ($A'=13, A=12$) $IS$ on various energy differences and compares them to previous works.

We have seen that the $S_{sms}$ parameter is strongly correlated to the energy, with negative and positive angular coefficients with respect to $p$ and $n$, respectively. The non-monotonous behavior of the SMS forbids us to generally conclude that any calculation similar to ours will result in upper or lower bounds to the differences in mass polarization expectation values. In our particular case however, Tables~\ref{tab7:resNR_C} and~\ref{tab7:resNR_C-} and Figure~\ref{fig7:conv} show that the convergence in $n$ is better achieved than in $p$ (truncated to $99.8\%$ or $99.3\%$). Therefore, we likely overestimate $S_{sms}$. Since we have globally $\Delta S_{sms} > \Delta E$, the estimations of the $IS$ on the detachment thresholds presented below, in particular on the $\EA$, are probably overly negative. Furthermore, we estimate that our non-relativistic values of $IS$ are reliable to about $2~10^{-1}$~m$^{-1}$ if the C$^-(^2D^o)$ is involved and of the order of $10^{-2}$~m$^{-1}$ if not.
Hence we cannot explain the disagreement between the isotope shift on the $\EA$ of Klopper \etal \cite{Kloetal:10a} and ours.
This is not an isolated discrepancy since we can extract from their results an (${18-16}$) $IS$ on the $\EA$ of oxygen of $-11.7$ m$^{-1}$ which is in disagreement with the experimental value of $-7.4(18)$~m$^{-1}$~\cite{Bloetal:01a}. Godefroid and Froese-Fischer~\cite{GodFro:99a} obtained an $IS$ of $-5.73$ m$^{-1}$ with a MCHF model, which is inside the experimental error bars (see also reference~\cite{Car:10a}). The MCHF approach has also proven its usefulness for the calculation of $IS$ on the $\EA$ of heavier systems~\cite{Caretal:10a,CarGod:11a}.

With regards to the difficulty to calculate the mass shifts, the discrepancy between the different calculations of the $IS$ on the neutral atom term separation, of the order of 0.1~m$^{-1}$, is understandable. From the comparison of our results and the ones of Kozlov \etal \cite{Kozetal:09a}, it is difficult to estimate an order of magnitude for the contribution of the relativistic effects. 

The neutral $^{3}P$ fine structure has been much more studied. It is known that the relativistic corrections to the mass shift operator are crucial when studying the isotope shift on the fine structure~\cite{Sto:61a,Sto:63a,Pal:87a,Sha:98a}. Veseth~\cite{Ves:85a} and more recently Kozlov \etal \cite{Kozetal:09a} performed calculations of the relativistic mass shifts in the $^3P$ multiplet of carbon, the first by treating perturbatively the fine-structure and nucleus-mass dependent Hamiltonians up to the third order, the second by using an all-electron CI method on the Dirac-Breit Hamiltonian and calculating the expectation value of the relativistic MS operator valid to the second order in $\alpha Z$. We easily estimate the relativistic corrections to the specific mass shift operator for the transition $^3P_1-{^3P}_2$ by comparing the equation (8) of Veseth to the Breit-Pauli fine-structure operator~\footnote{From reference~\cite{Ves:85a}, we find that the relativistic corrections to the $IS$ on the fine structure can be estimated using the Table VI of Veseth's paper as \mbox{$2/3[(8d) + (8e)]+(8f)$}.}. We obtain a correction of $+0.0375$ m$^{-1}$ for the $^{13-12}IS(^3P_1 - {^3P_2})$ which, combined with our result of Table~\ref{tab7:IS}, gives a total shift of $+0.014$~m$^{-1}$. This value is in good agreement with the observation and with Veseth's results. However, neither this observation nor the stability of the corrections of Table~\ref{tab7:rsms} demonstrates that the scalar relativistic effects on the $IS$ are reliable.

\begin{table*}
\caption{\label{tab7:transC}Einstein $A$ coefficients, in s$^{-1}$, for the $2p^2$ intra-configuration M1 and E2 transitions of carbon calculated using the BPCI wave functions based on the MR$_{98}$-I models for the active space $\lceil 10k \rceil$ and $\lceil 12k \rceil$, compared with the calculations of~\cite{Fro:06b}. Transitions vacuum wavelengths~($\lambda$) are reported in {\AA}ngstr\"om.}
\begin{ruledtabular}
\begin{tabular}{D{-}{-}{-1}llD{.}{.}{7}lD{.}{.}{7}lD{.}{.}{7}l}
 & & \multicolumn{4}{c}{This work} &\\
\cline{3-6}
& &  \multicolumn{2}{c}{$n=10$} & \multicolumn{2}{c}{$n=12$} & \multicolumn{2}{c}{Froese-Fischer~\cite{Fro:06b}} & \multicolumn{1}{c}{NIST~\cite{nist:levels}} \\
\cline{3-4}\cline{5-6}\cline{7-8}\cline{9-9}
\multicolumn{1}{c}{states}& type  & \multicolumn{1}{c}{$\lambda$} & \multicolumn{1}{c}{$A_{ki}$} & \multicolumn{1}{c}{$\lambda$} & \multicolumn{1}{c}{$A_{ki}$} &  \multicolumn{1}{c}{$\lambda$} & \multicolumn{1}{c}{$A_{ki}$}  & \multicolumn{1}{c}{$\lambda$} \\
\hline
^3P_0 \ -\ ^3P_1 & M1 & $6072(+3)$ & 8.033(-8)  & $6070(+3)$ & 8.041(-8) & $6052(+3)$ & 8.114(-8)  & $6090(+3)$ \\
^3P_1 \ -\ ^3P_2 & M1 & $3703(+3)$ & 2.656(-7)  & $3701(+3)$ & 2.660(-7) & $3700(+3)$ & 2.662(-7)  & $3704(+3)$ \\
                 & E2 &            & 3.529(-15) &            & 3.536(-15)&            & 3.633(-15) &  \\
^3P_0 \ -\ ^3P_2 & E2 & $2300(+3)$ & 1.696(-14) & $2299(+3)$ & 1.700(-14)& $2296(+3)$ & 1.754(-14) & $2304(+3)$ \\
^3P_2 \ -\ ^1D_2 & M1 & $9805$ & 2.370(-4) & $9814$ & 2.358(-4) & $9735$      & 2.245(-4)  & $9853$ \\
                 & E2 &        & 1.210(-6) &        & 1.209(-6) &             & 1.140(-6)  &  \\
^3P_1 \ -\ ^1D_2 & M1 & $9779$ & 7.513(-5) & $9788$ & 7.504(-5) & $9710$      & 7.544(-5)  & $9827$ \\
                 & E2 &        & 1.050(-7) &        & 1.037(-7) &             & 1.576(-7)  &  \\
^3P_0 \ -\ ^1D_2 & E2 & $9763$ & 7.789(-8) & $9772$ & 7.836(-8) & $9694$      & 6.242(-8)  & $9811$ \\
^3P_2 \ -\ ^1S_0 & E2 & $4604$ & 2.138(-5) & $4606$ & 2.130(-5) & $4587$      & 2.250(-5)  & $4629$ \\
^3P_1 \ -\ ^1S_0 & M1 & $4598$ & 2.368(-3) & $4600$ & 2.367(-3) & $4581$      & 2.381(-3)  & $4623$ \\
^1D_2 \ -\ ^1S_0 & E2 & $8679$ & 6.148(-1) & $8681$ & 6.135(-1) &             &            & $8730$ \\
\end{tabular}
\end{ruledtabular}
\end{table*}

\subsection{Transition probabilities}\label{sec7:trans}

We study the M1 and E2 transition probabilities between the $LSJ$ levels of the ground configurations of C and C$^-$. 
With the exception of $^1D_2 - {^1S_0}$ and the transitions between states of a same multiplet, the calculated Einstein coefficients are only non-zero thanks to the $LS$ relativistic mixing. The M1 channel of the $^4S^o_{3/2} -\ ^2D^o_{5/2}$ transition is itself only opened by $LS$ mixing of correlation CSFs.
The non-relativistic magnetic dipole and electric quadrupole interactions are computed using the BPCI wave functions based on the MR$_{98}$-I model described in the end of Section~\ref{sec7:bpci}. For the calculation of the Einstein $A$ transition rates between states that are developed in non-orthogonal orbital sets, we use the \textsc{biotr} program that is part of the ATSP2K package~\cite{Froetal:07a}.

The results of our calculations on the neutral atom are summarized in Table~\ref{tab7:transC}. The convergence of the $A$ coefficients with the active set is well achieved and the comparison with  the results of Froese-Fischer~\cite{Fro:06b} is favorable. Still, we can point out that, for the weak E2 transition rates of inter-term transitions  ($^3P_0 \ -\ ^1D_2$ and $^3P_1 \ -\ ^1D_2$), the relative change between the results of Froese-Fischer and ours is quite large.


The results for the C$^-$ are displayed in Table~\ref{tab7:transC-}.
There is other value available in the literature. To fill this gap, we compare our transition probabilities with others for the first elements of its iso-electronic sequence, \ie N~I and O~II. 
\emph{A priori}, the omission of the relativistic corrections to the M1 transition operator~\cite{Dra:71a,NemGod:09a} could be a serious limitation of our calculations. Indeed, Eissner and Zeippen~\cite{EisZei:81a} showed that for transitions between terms of the $2p^3$ configuration,  in particular, the relativistic corrections are of the same order of magnitude as the usual non relativistic amplitude (see N I and O II data of Table~\ref{tab7:transC-}).

However, as can be seen from Table~\ref{tab7:transC-} and in reference~\cite{Becetal:89a}, if the M1 channel becomes rapidly dominant with increasing $Z$ along the nitrogen-like sequence, the E2 channel remains, for low $Z$, of the same order of magnitude as the M1 channel. Furthermore, in the case of the C$^-$ system, the relativistic mixing of the $^2D^o_{3/2}$ and $^4S^o_{3/2}$ is even further suppressed by the diffuse nature of the $^2D^o$ state. The M1 channel is then only a small correction to the total $A$ in the C$^-(^4S^o-\ ^2D^o)$ transition, being of the same order of magnitude as the uncertainty with respect to the convergence with the active set ($\lceil 8k \rceil - \lceil 7i \rceil$).

\begin{table*}
\caption{Einstein $A$ coefficients, in s$^{-1}$, for the intra-configuration M1 and E2 transitions of all C$^-$ bound states, calculated using the BPCI wave functions based on the MR$_{98}$-I models for the active space $\lceil 4f \rceil$ to $\lceil 8k \rceil$. We compare these results with the corresponding nitrogen and nitrogen-like oxygen $A$ coefficients.
Transitions vacuum wavelengths~($\lambda$) are reported in {\AA}ngstr\"om.
The final set is obtained by re-normalizing the transition probabilities $\lceil 8k \rceil$ with these experimental ($^4S^o-^2D^o)$ and calculated ($^2D^o_{3/2-5/2}$) energy differences. \label{tab7:transC-}}
\begin{ruledtabular}
\begin{tabular}{lrD{.}{.}{7}D{.}{.}{7}rD{.}{.}{7}D{.}{.}{7}rD{.}{.}{8}D{.}{.}{8}}
 & \multicolumn{3}{c}{$^4S^o_{3/2} \ -\ ^2D^o_{3/2}$} & \multicolumn{3}{c}{$^4S^o_{3/2} \ -\ ^2D^o_{5/2}$} & \multicolumn{3}{c}{$^2D^o_{3/2} \ -\ ^2D^o_{5/2}$} \\
\cline{2-4}\cline{5-7}\cline{8-10}
&  \multicolumn{1}{c}{$\lambda$} &  \multicolumn{1}{c}{M1} &  \multicolumn{1}{c}{E2} & \multicolumn{1}{c}{$\lambda$} &  \multicolumn{1}{c}{M1} &  \multicolumn{1}{c}{E2} & \multicolumn{1}{c}{$\lambda$} &  \multicolumn{1}{c}{M1} &  \multicolumn{1}{c}{E2}\\
\hline
&\multicolumn{9}{c}{C$^-$ (Z $=6$)}\\
\raisebox{3ex}{}%
$n=4 $& $8813$ & 1.177(-6) & 7.628(-7) & $8815$ & 4.823(-8) & 1.118(-6) & $-5716(+4)$ & 8.664(-11) & 3.108(-22) \\
$n=5 $& $9356$ & 1.130(-6) & 8.915(-7) & $9357$ & 4.197(-8) & 1.297(-6) & $-1245(+5)$ & 8.415(-12) & 3.912(-23)\\ 
$n=6 $& $9590$ & 1.126(-6) & 1.019(-6) & $9590$ & 3.999(-8) & 1.470(-6) & $-6509(+5)$ & 5.893(-14) & 2.141(-26) \\ 
$n=7 $& $9801$ & 1.123(-6) & 1.125(-6) & $9800$ & 3.802(-8) & 1.605(-6) & $ 1519(+5)$ & 3.092(-12) & 4.209(-23) \\
$n=8 $& $9930$ & 1.080(-6) & 1.192(-6) & $9929$ & 3.663(-8) & 1.688(-6) & $ 9908(+4)$ & 1.109(-11) & 5.206(-22)\\
\raisebox{6ex}{}%
final && 1.030(-6) & 1.102(-6) && 3.493(-8) & 1.559(-6) && 5.771(-11) \\
\hline
&\multicolumn{9}{c}{N (Z $=7$)}\\
\raisebox{3ex}{}%
FFT\footnote{Reference \cite{FroTac:04a}.}   & $5199$ & 1.595(-5) & 4.341(-6) & $5202$ & 9.710(-7) & 6.595(-6) & $-1148(+4)$ & 1.071(-8) & \\
BZ nr\footnote{Reference \cite{ButZei:84a}, ``nr" means the non-relativistic M1 operator is used.} &        & 1.716(-5) & 3.822(-6) &        & 1.046(-6) & 5.880(-6) & $-1085(+4)$ & 1.239(-8) & \\
BZ$^\text{b}$  &        & 1.896(-5) &           &        & 2.445(-7) &           &              & 1.239(-8) \\
\raisebox{3ex}{}%
&\multicolumn{9}{c}{O$^+$ (Z $=8$)}\\
\raisebox{3ex}{}%
FFT$^\text{b}$   & $3727$ & 1.414(-4) & 2.209(-5) & $3730$ & 7.416(-6) & 3.382(-5) & $-5071(+3)$ & 1.241(-7) &  \\
Z87 nr\footnote{Reference \cite{Zei:87a}, ``nr" means the non-relativistic M1 operator is used.} &        & 1.45(-4)  & 2.13(-5)  &        & 7.58(-6)  & 3.30(-5)  & $-5119(+3)$ & 1.30(-7)  & \\
Z87$^\text{c}$  &        & 1.58(-4)  &           &        & 2.00(-6)  &           &              & 1.30(-7)
\end{tabular}
\end{ruledtabular}
\end{table*}

The inversion of the fine structure splitting of the $^{2}D^o$ term in C$^-$ with increasing active set reflects the difficulty to calculate this quantity for half filled shell systems, particularly in highly correlated systems. From this regard, it is also interesting to note that this fine structure is ``normal", \ie not inverted as in the heavier iso-electronic systems. In other words, it tends to behave like a less-than-half-filled-shell system.  

The experimental $(^4S^o_{3/2} -\ ^2D^o_J)$ transition is at $10087(9)$~\AA, and the $^2D^o$ calculated fine structure that we recommend (see Table~\ref{tab7:endiff}) is $+5703~10^4$~\AA.
Table~\ref{tab7:transC-} presents the final $A$ coefficients calculated with the $\lceil 8k \rceil$ active set and renormalized by the experimental ($^4S^o-\ ^2D^o_{J}$) energy separation, and our theoretical value for the $^2D^o$ fine structure (see Table~\ref{tab7:endiff}).

\section{Conclusion}\label{sec7:concl}

We performed large scale MCHF-CI calculations of the energy levels belonging to the lowest configuration of neutral carbon and all bound states of C$^-$, including the fine structures, hyperfine structures and isotope shifts. In addition, we calculated all $M1$ and $E2$ transition rates between the studied $LSJ$ states.

The overall precision of the non-relativistic expectation values is estimated to be about $0.3-0.8\%$. However, this imprecision on the total energy and $S_{sms}$ leads to a larger uncertainty on the differential effects. To our knowledge, the obtained non-relativistic energies are the most accurate \emph{ab initio} values to date. We obtained an extrapolated energy for the C$(^3P)$ state of $-$37.84465~E$_h$ in good agreement with the semi-empirical ``exact" value of $-$37.8450~E$_h$~\cite{Chaetal:93a}.

We conducted a careful study of the relativistic corrections deduced from the comparison of relativistic CI calculations with the corresponding non-relativistic calculations. Even if the so-deduced corrections do permit a relevant theory versus experiment comparison and that the correlation effects still dominate our uncertainties in many cases, an estimation of the relativistic effects on a firmer basis should be performed in more accurate studies.
We also note that the C$^-$ negative ion is very little affected by relativity.

We check the experimental hyperfine structure of the neutral carbon by replacing the theoretical parameters used in the original papers~\cite{Habetal:64a,Woletal:70a} by our values. The resulting hyperfine constants and $^{11}$C nuclear magnetic moments are in good agreement with previous experimental and theoretical studies.

As far as the transitions probabilities calculations are concerned, we find a good agreement of our neutral carbon $A$ coefficients with the ones of the literature. For the C$^-$ intra-configuration transitions, we expect the relativistic corrections to the M1 operator to be less important than in higher $Z$ iso-electronic systems. Once more, the missing correlation effects are equally limiting.

We find that the parametrization of the model in terms of the number of correlation layers ($\sim n$) and percentage of the wave function accounted by the MR in subsequent CI calculations ($=p$), provides useful tools for including a fixed percentage of the total correlation effects. It allows to establish lower bounds on detachment thresholds and, for calculations that are sufficiently converged with respect to $n$, upper bounds on $\Delta S_{sms}$ \mbox{(lower bound on $S_{sms}$)}.

\section*{Acknowledgements}
\noindent
TC is grateful to the ``\emph{Fonds pour la formation \`a la Recherche dans l'Industrie et dans l'Agriculture}" of Belgium for a Ph.D. grant (\emph{Boursier F.R.S.-FNRS}).
MRG and TC thank the \emph{Communaut\'e fran{\c c}aise of Belgium} (\emph{Action de Recherche Concert\'ee}) for financial support. MRG also acknowledges the Belgian National Fund for Scientific Research (FRFC/IISN Convention). Finally, the authors thank  Nathalie Vaeck and Per J\"onsson for fruitful discussions.
\end{document}